\documentclass[11pt,preprintnumbers,nofootinbib,prd]{revtex4-1}
\usepackage{graphicx,amsfonts,amsmath,amssymb,epsfig,color,mathrsfs,epsf}
\ifx\plotpoint\undefined\newsavebox{\plotpoint}\fi
\sbox{\plotpoint}{\rule[-0.200pt]{0.400pt}{0.400pt}}
\newcommand{\be}{\begin{equation}}
\newcommand{\ee}{\end{equation}}
\newcommand{\ba}{\begin{eqnarray}}
\newcommand{\ea}{\end{eqnarray}}
\newcommand{\nn}{\nonumber}

\begin{document}
\begin{flushright}
IC-2010/nnn
\end{flushright}\vspace{1cm}

\title{\centerline{Supersymmetry Breaking, Moduli Stabilization and Hidden $U(1)$ Breaking in $M$-Theory}}
\author{Bobby S. Acharya}\author{Mahdi Torabian}\email{mahdi@ictp.it}
\affiliation{International Centre for Theoretical
Physics, Strada Costiera 11, Trieste, Italy}

\begin{abstract}
We calculate and explore the moduli potential for $M$ theory compactified on $G_2$-manifolds in which the superpotential is
dominated by a single membrane instanton term plus one from an asymptotically free hidden sector gauge interaction. We show that
all moduli can be stabilized and that hidden sector gauge symmetries can be Higgsed at a high scale. We then compute the
spectrum of superpartner masses at the GUT scale and evolve it to the electroweak scale. We find a spectrum which is very
similar to the $G_2$-MSSM with light gauginos - accessible at the LHC - and a neutral wino dark matter candidate.
\end{abstract}
\maketitle


\section{Introduction}

String/$M$ theory vacua with four large spacetime dimensions are effectively described by a low energy
theory which typically has non-Abelian gauge symmetry, chiral fermions and hierarchical Yukawa couplings.
These facts were essentially evident in 1984 \cite{Candelas-etal}. Since these are some of the key ingredients of the Standard Model
of particle physics, this provides some essential motivation for pursuing questions in so-called "string phenomenology".

At that time, the vacua mainly under consideration were exactly supersymmetric - a condition motivated
by the relevance of supersymmetry to the stability of the mass scale of the Standard Model under quantum
corrections. The problem of how supersymmetry is broken is closely related to the moduli problem(s).
Supersymmetric string/$M$ theory vacua have moduli fields --  
namely, the extra dimensions have zero modes in the supersymmetric limit. These zero modes are continuous parameters
of the metric on the extra dimensions (e.g. Calabi-Yau or $G_2$-holonomy metrics). They parametrize
the size and shape of the extra dimensions and appear in the low energy theory as fields with Planck suppressed
interactions with matter. Having massless scalars with Newtonian couplings is problematic phenomenologically
for a number of reasons such as "fifth force experiments" and cosmological constraints.

The standard lore was that, once supersymmetry breaking is "understood", these moduli would receive masses from
supersymmetry breaking effects. Since all couplings and masses are functions of the moduli fields in string/$M$ theory
any supersymmetry breaking effects in the effective Lagrangian will depend upon them.
Therefore, whatever mechanism is responsible for supersymmetry breaking will {\it necessarily} also generate a contribution
to the moduli potential. If supersymmetry is the solution to the hierarchy problem, then such contributions
to the moduli potential must have a size of order $M_{\rm susy}^4$ (where the value of $M_{\rm susy}$ depends upon
how supersymmetry breaking is "mediated" to the observable sector). In any realistic case, $M_{\rm susy}$ is orders
of magnitude smaller than, say, the Planck scale. One of the simplest ideas for generating a small scale of supersymmetry
breaking, which goes back to Witten \cite{Witten-81} and others is that this small scale is the strong dynamics
scale of a new hidden sector interaction, in close analogy with the dynamically generated scale of QCD. 

Putting all of these ideas together, strong dynamics in a new hidden sector interaction should i) generate 
the hierarchy between $m_{pl}$ and $M_{W}$, ii) break supersymmetry and iii) generate a potential which stabilizes all
the moduli fields. Attempts at realizing these ideas in string theory have met with partial success over the years.
Nilles \cite{Nilles:1982ik} made the earliest investigations in the supergravity context, \cite{Casas-Munoz-Ross, Derendinger:1985kk} considered
the heterotic string. In all of the investigations in string theory, the problem could only be
considered to be partly solved in the sense that it was not understood how to stabilize ALL of the moduli fields whilst simultaneously
generating the mass hierarchy.

The problem was finally given a fairly satisfactory solution \cite{G2-MSSM} in the context of $M$ theory vacua in which the seven extra dimensions
form a manifold of $G_2$-holonomy \cite{Acharya-Gukov,Acharya}. There is a simple explanation for this success: in $M$ theory, in the absence of fluxes, all the moduli fields reside in superfields which undergo shift symmetries (because they all contain axions). This implies that only "non-perturbative" contributions to the potential are
allowed -- such as strong dynamics of the type we are interested in for the hierarchy problem. Thus, the potential energy will always be exponentially small in $M$ theory. This is not the case in the heterotic string and other string theories. For instance, there are no symmetries preventing perturbative contributions to the potential for complex structure moduli in the heterotic or Type IIB string theories.

In \cite{G2-MSSM} the main idea was to consider hidden sectors which consist of two or more asymptotically free gauge sectors\footnote{
One hidden sector gauge group is actually enough to stabilize all the moduli, however the vacuum is not in a region where the supergravity approximation
is valid.}. 
The purpose of this paper is to consider the effect and role of other types of terms in the superpotential on the moduli potential.
In $M$ theory all terms in the potential are non-perturbative. For instance an ordinary cubic Yukawa coupling between three charged chiral superfields is generated by a membrane instanton
which wraps a 3-dimensional submanifold of the extra dimensions which passes through the points where the fields are localized \cite{Acharya-Witten}. As such it will be exponentially suppressed by the volume of this cycle. 
We will consider a 
hidden sector $U(1)$ gauge group with charged matter with a cubic Yukawa coupling as well as a single asymptotically free hidden sector gauge group (which
turns out to be necessary for obtaining vacua treatable within the supergravity approximation). 

We will follow the ideas of \cite{G2-MSSM} to understand the minima of the supergravity potential. We find that all moduli can indeed be stabilized by
these contributions. We then go on to calculate the observable sector supersymmetric partner masses and couplings at the electroweak scale.
We will see that this extends the framework of \cite{G2-MSSM} in the sense that it permits for a wider range of gravitino masses (i.e. supersymmetry breaking scales), however the low energy phenomenological features for observationally consistent $G_2$-manifolds remain essentially the same as those in the \cite{G2-MSSM}.
The discussion section at the end of the paper briefly summarizes the main phenomenological features and suggests some further directions to investigate.
 
\section{Four Dimensional Effective Theory and Moduli Stabilization}
The complexified moduli space ${\cal M}_{b_3}$ of $M$ theory compactified on a real manifold manifold $X_7$ of $G_2$ holonomy is parametrized by holomorphic coordinates 
\be z_i = t_i + is_i \ ,\ee
where $i$ runs from 1 to $b_3$, the $3^{rd}$ Betti number of $X_7$. Moduli and axion fields, respectively denoted by $s_i$ and $t_i$, arise from the 11 dimensional metric and 3-form field as Kaluza-Klein zero modes of $X_7$. These fields pair up to form the bosonic components of chiral superfields in four dimensions.\footnote{The zero modes of the eleven dimensional gravitino field along $X_7$ gives rise to the four dimensional gravitino and modulino-axino fields required
by supersymmetry.} All the parameters of the low scale physics are functions of the moduli vacuum expectation values, so stabilizing the moduli is a fundamental step towards phenomenological
implications. As an side, note that the axions, being periodic fields, are not moduli.

\subsection{Moduli Stabilization and Symmetry Breaking}

The tree level K\"ahler potential of the effective four dimensional supergravity theory is \cite{Beasley:2002db} 
\be\label{Kahler-potential} K = -3\ln\big(4\pi^{1/3}V_X\big) \ ,\ee where $V_X$ is the volume of $X_7$ in units of the eleven dimensional Planck length. 
$V_X$ is a homogeneous function of moduli of degree 7/3. It has been demonstrated in \cite{Kahler-Independence} that the observable sector matter spectrum (e.g. superpartner masses) depends very weakly on the detailed form of the moduli K\"ahler potential - a fact due to the homogeneity of $V_X$.

We will consider $G_2$ manifolds which give rise to i) a hidden sector gauge theory which confines at low energies ii) a hidden sector $U(1)$ gauge group
under which there are three charged matter superfields $T_a$ with a cubic superpotential. The confining gauge theory arises from a 
co-dimension four orbifold singularity along a 3-dimensional sub-manifold of $X_7$. Due to the axionic shift symmetry and holomorphy, the
coupling of the cubic term in $W$ is generated by membrane instantons and is therefore exponentially suppressed by moduli vacuum expectation values.

The total superpotential is thus given by 
\be W = A\,e^{2\pi if/P} + \lambda\,T_1T_2T_3\,e^{2\pi i\tilde f}\ .\ee
In the above $A$ (due to PQ shift symmetry) and $\lambda$ are ${\cal O}(1)$ constants. The constant $P$ is the beta function coefficient of the confining gauge theory.
These two terms are regarded as the leading contributions to $W$. The superpotential can, in principle contain many additional non-perturbative contributions if $X_7$ contains additional supersymmetric 3-cycles. 

The gauge kinetic function of the gauge theory in the hidden sector is a linear combination of the moduli and given by 
\be f = \sum_{i=1}^{b_X^3}N_iz_i = \frac{\theta}{2\pi} + i\alpha_{\rm HS}^{-1}\ .\ee 
The imaginary part is the inverse of the hidden sector gauge coupling $\alpha_{\rm HS}^{-1}$ which plays a central role in this work. Since, in the UV 
hidden sector is weakly coupled, we can regard the coupling 
as an expansion parameter in the eventual vacuum we will study. This is consistent as long as the vacuum has moduli vacuum expectation values such that $\alpha_{\rm HS} \ll1$.
$\tilde f$ is also a linear combination of moduli 
\be \tilde f = \sum_{i=1}^{b_X^3}\tilde N_iz_i\ ,\ee 
the imaginary part of which is the volume of the 3-cycle wrapped by the membrane instanton. $N_i$ and $\tilde N_i$ are positive integers which are determined by the homology classes of the 3-cycles.  These are microscopic parameters determined by the details of the compactification manifolds. 

The K\"ahler potential for the matter fields in the $U(1)$ sector is given by 
\be\label{matter-Kahler-potential} K^{U_1} = \sum_{a=1}^3\frac{\overline T_aT_a}{V_X}\ .\ee This moduli dependent form has been proposed in different ways for calculating the moduli dependence of the matter kinetic terms \cite{Kahler-Independence}. 

The scalar potential in 4-dimenasional ${\cal N}=1$  SUGRA is determined by 
\be\label{scalar-potential} V =
e^G\Big(\sum_IG^IG_I - 3\Big)\ .\ee 
The summation is over all the moduli and matter fields ($I=i,a$). The K\"ahler function $G$ is defined as 
\be G=K + \ln|W|^2\ .\ee
The $F$-terms are given by K\"ahler covariant derivatives \be G_I = \frac{{\cal D}_IW}{W} = \frac{\partial_IW}{W} + \partial_IK\ ,\ee  and $G^I = \sum_{\bar J}K^{I\bar J}G_{\bar J}$ where $K^{I\bar J}$ is the inverse K\"ahler metric. 

Whereas there are two terms in the superpotential, as a cross term, only one linear combination of axions $t_i$ and phases of the complex scalar fields $\tau_a$ appears in the scalar potential. Minimizing the scalar potential fixes this combination to be (for $A>0$) 
\be \cos\Big(\sum_iw_iN_it_i-\tau_1-\tau_2-\tau_3\Big)=-1\ ,\ee 
where we have defined $w_i$ as \be w_i = \frac{2\pi}{P} - \frac{2\pi}{N_i/\tilde N_i}\ .\ee 
In order to stabilize the rest of the axions, we need to consider some $b_3+2$ more subleading terms in the superpotential generated by
membrane instantons (obviously with larger actions than the term considered above). The extra cross terms generate a potential stabilizing the
remaining axions and is studied in detail in \cite{axiverse}.

In order to simplify the study we introduce the following notation
\ba \label{x} x &=& \frac{\lambda}{A}\,T_{01}T_{02}T_{03}\,e^{\sum_i w_iN_is_i}\ .\ea 
The superpotential can be rewritten more compactly as
\be W = A(1-x)\ e^{-2\pi N\cdot s/P}e^{2\pi iN\cdot t/P}\ .\ee
The $F$-terms can be derived explicitly as follows 
\ba G_i &=& iN_i\frac{y_i}{1-x}\bigg(1 + \frac{3}{2}\frac{a_i}{N_is_i}\frac{1-x}{y_i}\bigg(1+\frac{\sum_a\overline T_aT_a}{3V_X}\bigg)\bigg)\ ,\\ G_{T_a}  &=& \frac{\overline T_a}{V_X}\bigg(1-\frac{x}{1-x}\frac{V_X}{\overline T_aT_a}\bigg)\ ,\ea 
where 
\ba\label{y_i} y_i &=& \frac{2\pi}{P} - x\frac{2\pi}{N_i/\tilde N_i}\ .\ea 
In the above and in the following, there is no summation over repeated indices $i$ and $a$ unless explicitly written. We have also introduced some variables $a_i$ defined by 
\be\label{a_i-def} a_i \equiv \frac{\partial\ln V_X}{\partial\ln s_i} \ ,\ee 
such that $\sum_i a_i=7/3$. We then can write
\be \frac{\partial K}{\partial z_i} = \frac{1}{2i}\frac{\partial K}{\partial s_i} = -\frac{3}{2i}\frac{\partial\ln V_X}{\partial s_i} \equiv -\frac{3}{2i}\frac{a_i}{s_i}\ .\ee

Using the K\"ahler potential \eqref{Kahler-potential}, one can obtain the components of the K\"ahler metric. Using the orthonormality condition 
\be \sum_{\bar J} K^{I\!\bar J}K_{\bar JK} = \delta^I_K \ ,\ee
 one can read the non-zero components of the inverse K\"ahler metric.  After some work, one obtains the following expression for the raised index $F$-terms \ba G^i &=& -is_i\bigg(\frac{4}{3}\frac{1}{a_i}\sum_{\bar j}\frac{y_{\bar j}}{1-x}s_{\bar j}N_{\bar j}(\Delta^{-1})^{i\bar j} + 2\frac{x}{1-x} + 2\bigg)\frac{1}{1+\frac{\sum_a\overline T_aT_a}{3V_X}}\ , \\\!\!\! G^{T_a}  &=& -T_a\bigg(\frac{4}{3} + \frac{x}{1-x}\frac{V_X}{\overline T_aT_a} + \frac{2}{3}\frac{1}{1+\frac{\sum_a\overline T_aT_a}{3V_X}}\sum_{i}\frac{y_i}{1-x}s_iN_i - \frac{7}{9}\frac{1}{1+\frac{\sum_a\overline T_aT_a}{3V_X}}\frac{\overline T_aT_a}{V_X} + \frac{7}{9}\frac{1}{1+\frac{\sum_a\overline T_aT_a}{3V_X}}\frac{x}{1-x}\bigg).\nn\\\ea 
 In the above, the matrix $\Delta_{ij}$ is defined as \be \Delta_{ij} \equiv \frac{1}{\frac{3a_j}{s_is_j}}\hat K_{ij} = \delta^{ij} + \frac{1}{a_i}\Big(\frac{1}{3}\delta_{ij}K_is_j+\frac{1}{3}K_{ij}s_is_j\Big)\ , \ee where the matrix $\hat K_{ij}$ is the Hessian of $\hat K$ which is related to K\"ahler metric by a factor of 4. One can easily check that the matrix $\Delta_{ij}$ satisfies the following properties \be \sum_{i,j=1}^{b_X^3}\Delta_{ij} = 1 \qquad,\qquad \sum_{j=1}^{b_X^3}\Delta_{ij}a_j = a_i \qquad,\qquad \sum_{j=1}^{b_X^3}\big(\Delta^{-1}\big)^{ij}\Delta_{ij} = \delta^i_j\ .\ee 
  Finally, putting all the pieces together one obtains the following expression for the scalar potential
\be\label{potential}\begin{split} V(s_i,T_a) = \frac{e^{\sum_a\overline T_aT_a/V_X}}{64\pi V_X^3}\ A^2(1-x)^2e^{-4\pi/P\sum_i N_is_i}\Bigg[& \frac{4}{3}\frac{1}{1+\frac{\sum_a\overline T_aT_a}{3V_X}}\sum_{i}\sum_{\bar j}\frac{1}{a_i}N_is_i\frac{y_i}{1-x}\big(\Delta^{-1}\big)^{i\bar j} N_{\bar j}s_{\bar j}\frac{y_{\bar j}}{1-x} \cr +& 4\frac{1}{1-x}\frac{1}{1+\frac{\sum_a\overline T_aT_a}{3V_X}}\sum_iN_is_i\frac{y_i}{1-x} \cr +& \frac{x^2}{(1-x)^2}\bigg(\sum_a\frac{V_X}{\overline T_aT_a} + \frac{7}{3}\frac{1}{1+\frac{\sum_a\overline T_aT_a}{3V_X}} \bigg) \cr +& \frac{x}{1-x}\bigg(8 - \frac{14}{9}\frac{1}{1+\frac{\sum_a\overline T_aT_a}{3V_X}}\sum_a\frac{\overline T_aT_a}{V_X} \bigg) \cr -& \frac{4}{3}\sum_a\frac{\overline T_aT_a}{V_X} + \frac{7}{9}\frac{1}{1+\frac{\sum_a\overline T_aT_a}{3V_X}}\sum_a\left(\frac{\overline T_aT_a}{V_X}\right)^2 + 4\Bigg]\ .\end{split}\ee

In order to study the vacua of the potential, we make the following ansatz for the moduli vacuum expectation values at the minimum  
\be\label{moduli-ansatz} s_i = \frac{a_i}{N_i}\,\frac{1-x}{y_i}\,s\ .\ee 
The unknowns $a_i$ and $s$ will be determined later. Then, the inverse coupling of the hidden sector gauge theory is given by 
\be\label{alpha-hs-ansatz} \alpha_{\rm HS}^{-1} = (1-x)s\sum_{i=1}^{b_X^3}\frac{a_i}{y_i}\ .\ee  
In order to study the minima of the potential, we use the fact that the hidden sector gauge theory in the UV is weakly coupled 
\be \alpha_{\rm HS}\ll 1\ .\ee 
We keep working in this regime and regard the gauge coupling $\alpha_{\rm HS}$ as our expansion parameter and expand our solutions in its powers. We understand from \eqref{alpha-hs-ansatz} that the weak coupling limit is attained when one or more of the $y_i$, $1\leq i \leq b_X^3$, are small
\be y_i\sim \alpha_{\rm HS} \rightarrow 0,\qquad i\in \chi \ .\ee
For the moment, we also assume that $s$ is non-zero and finite in this limit, a crucial assumption to be verified later. For simplicity, we use the following parametrization 
\be y_i = -\zeta_i\, \alpha_{\rm HS} = -\zeta\, \alpha_{\rm HS} + (\zeta-\zeta_i)\alpha_{\rm HS} 
,\qquad i\in \chi,\ee 
where $\zeta$ and $\zeta_i$ are ${\cal O}(1)$ constants. From the definition of $y_i$ \eqref{y_i} and by the following expansion
\be \frac{N_i}{\tilde N_i} = n + n_i \alpha_{\rm HS}+{\cal O}(\alpha_{\rm HS}^2),\qquad i\in \chi\ ,\ee
we find that
\be y \equiv -\zeta\,\alpha_{\rm HS} = \frac{2\pi}{P}-x\frac{2\pi}{n}\ ,\ee
and 
\be \zeta - \zeta_i = \frac{2\pi}{P}\frac{n_i}{n},\qquad i\in \chi\ .\ee
The constant $\zeta$ is a measure of the relative difference between the two homology classes which generate the two terms in the superpotential.
One can show that it is less than $7/2$ (for the case all of the $y_i$ go to zero). Furthermore, in this parametrization we can also write $w_i$ as follows: \be w_i = w + w'_i \alpha_{\rm HS} + {\cal O}(\alpha_{\rm HS}^2)  \equiv \frac{2\pi}{P} - \frac{2\pi}{n} + (\zeta-\zeta_i)\frac{P}{n}\alpha_{\rm HS} + {\cal O}(\alpha_{\rm HS}^2),\qquad i\in \chi .\ee
Finally, from \eqref{y_i} we can show that 
\be\label{x-0th} x = \frac{n}{P} \Big(1+\zeta \frac{\alpha_{\rm HS}\cdot P}{2\pi}\Big) + {\cal O}(\alpha_{\rm HS}^2) \ .\ee 
The gauge coupling in the hidden sector can be re-expressed  as
\be \alpha_{\rm HS}^{-1} = \frac{1-x}{y}\,\zeta s \sum\frac{a_i}{\zeta_i} + (1-x)s\sum\frac{a_i}{y_i} \approx \frac{1-x}{y}\,\zeta s\sum\frac{a_i}{\zeta_i}\ .\ee
Thus, the ansatz for the moduli in the two sets can be rewritten as follows
\ba s_i &=& \frac{1}{N_i}\frac{a_i/\zeta_i}{\sum a_i/\zeta_i}\, \alpha_{\rm HS}^{-1}\ ,\qquad i\in \chi,\\ &=& \frac{1}{N_i}\frac{a_i/y_i}{\sum a_i/\zeta_i}\, y\alpha_{\rm HS}^{-1}\ ,\quad\ i\notin \chi, \ea 
Moreover, from \eqref{x} one can see that 
\be -\sum_i w_iN_is_i = \ln\left(\frac{1}{x}\frac{\lambda}{A}\ T_{01}T_{02}T_{03}\right)\ ,\ee 
which to the leading order in $\alpha_{\rm HS}$ can be written as
\be \alpha_{\rm HS}^{-1} \approx \frac{1}{2\pi/n-2\pi/P}\ \ln\left(\frac{P}{n}\frac{\lambda}{A}\ T_{01}T_{02}T_{03}\right)\ .\ee 
We take this as a constraint. For positivity of the coupling, either of the two following conditions have to be met 
\ba\label{branch-1} i&:&\quad \frac{n}{P}<1 \quad {\rm and}\quad \frac{n}{P}<\frac{\lambda}{A}T_{01}T_{02}T_{03}\ , \\\label{branch-2} ii&:&\quad \frac{n}{P}>1\quad {\rm and}\quad \frac{n}{P}>\frac{\lambda}{A}T_{01}T_{02}T_{03}\ . \ea

In order to determine the moduli vev's, in the first step one must determine the $a_i$. Putting back the ansatz \eqref{moduli-ansatz} into the definition of $a_i$ \eqref{a_i-def} one gets a system of $b_3$ transcendental equations which, in principle, completely fix the $a_i$ 
\ba a_i - \frac{\partial(-K/3)}{\partial\ln s_i} \bigg|_{s_i=\frac{1}{N_i}\frac{a_i/\zeta_i}{\sum a_i/\zeta_i}\, \alpha_{\rm HS}^{-1}} &=& 0\ , i\in \chi\ , \\ a_i - \frac{\partial(-K/3)}{\partial\ln s_i}\bigg|_{s_i=\frac{1}{N_i}\frac{a_i/y_i}{\sum a_i/\zeta_i}\, y\alpha_{\rm HS}^{-1}} &=& 0\ ,i\notin \chi\ .\ea 
For the moduli to be positive, for positive $N_i$, the solutions for $a_i$ must all be positive. One can numerically check that for quite generic sets of parameters the above system of equations yield positive solutions for $a_i$. 

Next we determine $s$ in the limit of interest, {\it i.e.} $y_i\rightarrow 0$, and verify the initial assumption. We minimize the scalar potential \eqref{scalar-potential} with respect to the moduli $s_i$ 
\be \frac{\partial V}{\partial s_i}\bigg |_{y\rightarrow 0} = 0\ .\ee 
We multiply each of these $b_3$ equations by $s_i$ and sum all of them together. This gives an equation for $s$ which can be used to determine $\alpha_{\rm HS}$ and this, in turn can determine the individual moduli vacuum expectation values from our ansatz \eqref{moduli-ansatz}
\be\begin{split} \bigg[&\frac{8}{3}\zeta\frac{w}{y}\, x s^2 - 4\frac{x}{1-x}+ 4\zeta\frac{w}{y}\frac{x}{1-x}\, s\bigg]\frac{1}{1+\frac{\sum_a\overline T_aT_a}{3V_X}} \cr -& 2\frac{x^2}{(1-x)^2} \bigg(\sum_a\frac{V_X}{\overline T_aT_a} + \frac{7}{3}\frac{1}{1+\frac{\sum_a\overline T_aT_a}{3V_X}}\bigg) - \frac{x}{1-x} \bigg(8-\frac{14}{9}\frac{1}{1+\frac{\sum_a\overline T_aT_a}{3V_X}}\sum_a\frac{\overline T_aT_a}{V_X} \bigg) = 0\ .\end{split}\ee 
Solving the above equation to the first subleading order in $y$ results in 
\be\label{s-solution} s = -\frac{3}{2}\frac{1}{1-x} + \left[\frac{1}{1-x} - \bigg(\frac{5}{14}+\frac{3}{14}\frac{x}{1-x}\sum_a\frac{V_X}{\overline T_aT_a}\bigg)\bigg(1+\frac{\sum_a \overline T_aT_a}{3V_X}\bigg)\right]\frac{y}{\zeta w}\ .\ee 
We can see that this solution is non-zero and finite when $y\rightarrow 0$ and so is self-consistent. This is the solution which describes the minimum of the potential. The second root which goes to zero in the limit $y\rightarrow 0$ corresponds to a maximum of the potential.

In the minimum, one reads the $F$-terms as follows
\ba G_i &=& iN_iy_i\bigg(\frac{1}{1-n/P} - \bigg(1+\frac{\sum_a\overline T_aT_a}{3V_X}\bigg)\bigg)\ ,\\ G_{T_a} &=& \frac{\overline T_a}{V_X}\bigg(1-\frac{n/P}{1-n/P}\Big(1+\zeta\frac{\alpha_{\rm HS}\cdot P}{2\pi}\Big)\frac{V_X}{\overline T_aT_a}\bigg)\ ,\ea 
and also 
\ba G^i &=& -is_i\bigg(\frac{4}{3}\frac{1}{1+\frac{\sum_a\overline T_aT_a}{3V_X}}\frac{1}{1-n/P}-\frac{10}{21}- \frac{2}{7}\frac{n/P}{1-n/P}\sum_a\frac{V_X}{\overline T_aT_a}\bigg)\frac{n/P}{1-n/P}\zeta\frac{\alpha_{\rm HS}\cdot P}{2\pi}\ , \\\!\!\! G^{T_a} &=& -T_a\bigg(\frac{4}{3} + \frac{n/P}{1-n/P}\frac{V_X}{\overline T_aT_a} - \frac{7}{9}\frac{1}{1+\frac{\sum_a\overline T_aT_a}{3V_X}}\frac{3-n/P}{1-n/P} - \frac{7}{9}\frac{1}{1+\frac{\sum_a\overline T_aT_a}{3V_X}}\frac{\overline T_aT_a}{V_X}\bigg)\\  &&-T_a\bigg(\frac{n/P}{1-n/P}\frac{V_X}{\overline T_aT_a}  - \frac{5}{9}\frac{1}{1+\frac{\sum_a\overline T_aT_a}{3V_X}}\frac{7}{1-n/P} + \frac{5}{9} + \frac{1}{3}\frac{n/P}{1-n/P}\sum_a\frac{V_X}{\overline T_aT_a} \cr &&\qquad\ + \frac{7}{9}\frac{n/P}{1-n/P}\frac{1}{1+\frac{\sum_a\overline T_aT_a}{3V_X}}\bigg)\frac{n/P}{1-n/P}\zeta\frac{\alpha_{\rm HS}\cdot P}{2\pi}\ .\nn\ea 
The expression for $s$ in the minimum is 
\be s = -\frac{3}{2}\frac{1}{1-n/P} - \bigg(\frac{5}{2}\frac{1}{1-n/P} - \Big(\frac{5}{14}+\frac{3}{14}\frac{n/P}{1-n/P}\sum_a\frac{V_X}{\overline T_aT_a}\Big)\Big(1+\frac{\sum_a\overline T_aT_a}{V_X}\Big)\bigg)\frac{n/P}{1-n/P}\zeta\frac{\alpha_{\rm HS}\cdot P}{2\pi}\ .\ee

The scalar potential at the minimum of the moduli to subleading order in $\alpha_{\rm HS}$ is read as follows \be\label{potential-at-minimum}\begin{split} V(T_a)\big |_{min} = \frac{A^2c^{-3}}{64\pi}&e^{\sum_a\overline T_aT_a/V_X} \alpha_{\rm HS}^{-7}e^{-4\pi/\alpha_{\rm HS}\cdot P}\times\cr\times\Bigg[4&(1-n/P)^2\Big(1-2\frac{n/P}{1-n/P}\zeta\frac{\alpha_{\rm HS}\cdot P}{2\pi}\Big) - 7\frac{1}{1+\frac{\sum_a\overline T_aT_a}{3V_X}} \cr +&(n/P)^2\Big(1+ 2\zeta\frac{\alpha_{\rm HS}\cdot P}{2\pi}\Big) \bigg(\sum_a\frac{V_X}{\overline T_aT_a} + \frac{7}{3}\frac{1}{1+\frac{\sum_a\overline T_aT_a}{3V_X}} \bigg) \cr +& n/P(1-n/P)\Big(1-2\frac{n/P}{1-n/P}\zeta\frac{\alpha_{\rm HS}\cdot P}{2\pi}\Big) \bigg(8 - \frac{14}{9}\frac{1}{1+\frac{\sum_a\overline T_aT_a}{3V_X}}\sum_a\frac{\overline T_aT_a}{V_X} \bigg) \cr -& \frac{4}{3}(1-n/P)^2\Big(1-2\frac{n/P}{1-n/P}\zeta\frac{\alpha_{\rm HS}\cdot P}{2\pi}\Big) \sum_a\frac{\overline T_aT_a}{V_X} \cr +& \frac{7}{9}(1-n/P)^2\Big(1-2\frac{n/P}{1-n/P}\zeta\frac{\alpha_{\rm HS}\cdot P}{2\pi}\Big) \frac{1}{1+\frac{\sum_a\overline T_aT_a}{3V_X}}\sum_a\left(\frac{\overline T_aT_a}{V_X}\right)^2 \Bigg]\ .\end{split}\ee 

\subsection{Supersymmetric Vacua}
We first consider the existence of supersymmetric vacua. The condition for a supersymmetric vacuum are vanishing $F$-terms
\ba G_i &=& 0\ , \\ G_{T_a} &=& 0\ ,\ea which implies
\ba s_i &=& -\frac{3}{2}\frac{a_i}{N_i}\frac{1}{y_i} = \frac{1}{2\pi}\frac{1}{N_i}\frac{a_i/y_i}{\sum_i a_i/y_i}\frac{P\ln \Big(\frac{P}{n}\frac{\lambda}{A}T_{01}T_{02}T_{03}\Big)}{P/n-1}\ ,\\
\label{TT-SUSY}\frac{\overline T_aT_a}{V_X} &=& \frac{x}{1-x} \approx \frac{n/P}{1-n/P}\Big(1+\zeta\frac{\alpha_{\rm HS}\cdot P}{2\pi}\Big) 
\ .\ea From above, for supersymmetric solutions to exist, it is necessary to satisfy \be\label{susy-condition} (n/P)<1\ . \ee For the moduli to be positive, the constants have to satisfy the conditions of the two different branches \eqref{branch-1} or \eqref{branch-2}. Thus, the SUSY condition \eqref{susy-condition} is consistent with branch $i$. 

\subsection{Metastable de Sitter Vacua}
Here we turn to non-supersymmetric minima. In order to find the vacua and the vev's of the $T_a$, we minimize \eqref{potential-at-minimum}  
\be\label{eq-1} \frac{\partial V|_{\rm min}\big(\overline T_aT_a/V_X,\alpha_{\rm HS}\big) }{\partial(\overline T_aT_a/V_X)} = 0\ .\ee 
Thus, we get a system of three equations \be\begin{split}&\big(1-n/P\big)^2\bigg(1-2\frac{n/P}{1-n/P}\frac{\alpha_{\rm HS}\cdot P}{2\pi}\bigg)\bigg(- \frac{4}{3} + \frac{7}{27}\frac{1}{1+\frac{\sum_a\overline T_aT_a}{3V_X}}\frac{\overline T_aT_a}{V_X}\bigg(6 - \frac{1}{1+\frac{\sum_a\overline T_aT_a}{3V_X}}\frac{\overline T_aT_a}{V_X}\bigg)\bigg) \cr & - n/P\big(1-n/P\big)\bigg(1+\frac{1-2n/P}{1-n/P}\frac{\alpha_{\rm HS}\cdot P}{2\pi}\bigg)\frac{14}{27}\frac{1}{1+\frac{\sum_a\overline T_aT_a}{3V_X}}\bigg(3 - \frac{1}{1+\frac{\sum_a\overline T_aT_a}{3V_X}}\frac{\overline T_aT_a}{V_X}\bigg) \cr & - (n/P)^2\bigg(1+\frac{2}{1-n/P}\frac{\alpha_{\rm HS}\cdot P}{2\pi}\bigg)\frac{V_X}{\overline T_aT_a}\bigg(\frac{V_X}{\overline T_aT_a} + \frac{7}{9}\frac{1}{1+\frac{\sum_a\overline T_aT_a}{3V_X}}\frac{\overline T_aT_a}{V_X} \bigg) + \frac{7}{3}\frac{1}{\Big(1+\frac{\sum_a\overline T_aT_a}{3V_X}\Big)^2} = 0\ .\end{split}\ee 
Furthermore, it should at least be possible to tune the cosmological constant to zero for phenomenological reasons
\be\label{eq-2} V|_{\rm min}\big(\overline T_aT_a/V_X,\alpha_{\rm HS}\big) = 0 \ ,\ee 
This gives the condition 
\be\begin{split}&\big(1-n/P\big)^2\bigg(1-2\frac{n/P}{1-n/P}\frac{\alpha_{\rm HS}\cdot P}{2\pi}\bigg)\bigg(4- \frac{4}{3}\sum_a\frac{\overline T_aT_a}{V_X} + \frac{7}{9}\frac{1}{1+\frac{\sum_a\overline T_aT_a}{3V_X}}\sum_a\left(\frac{\overline T_aT_a}{V_X}\right)^2\bigg) \cr &+ n/P\big(1-n/P\big)\bigg(1+\frac{1-2n/P}{1-n/P}\frac{\alpha_{\rm HS}\cdot P}{2\pi}\bigg)\bigg(8 - \frac{14}{9}\frac{1}{1+\frac{\sum_a\overline T_aT_a}{3V_X}}\sum_a\frac{\overline T_aT_a}{V_X}\bigg) \cr &+ (n/P)^2\bigg(1+\frac{2}{1-n/P}\frac{\alpha_{\rm HS}\cdot P}{2\pi}\bigg)\bigg(\sum_a\frac{V_X}{\overline T_aT_a} + \frac{7}{3}\frac{1}{1+\frac{\sum_a\overline T_aT_a}{3V_X}} \bigg) - 7\frac{1}{1+\frac{\sum_a\overline T_aT_a}{3V_X}} 
= 0\ .\end{split}\ee 

\begin{figure}[h]
\begin{center}$
\begin{array}{cc}
\hspace{-5mm}\includegraphics[scale=.8]{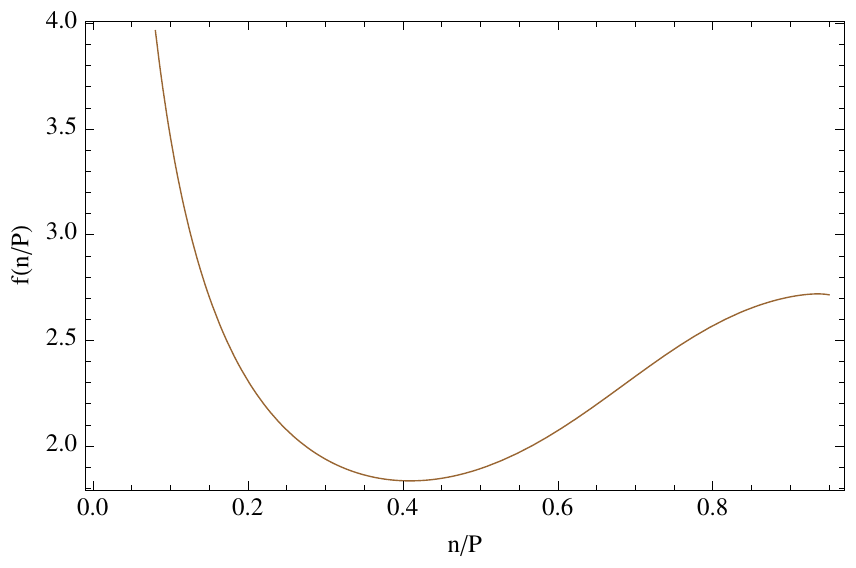} &
\includegraphics[scale=.8]{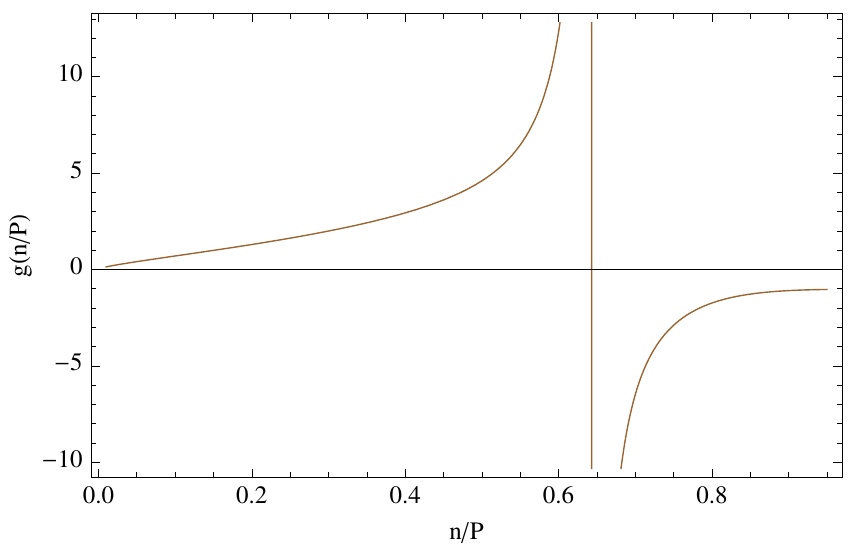}
\end{array}$
\end{center}
\vspace{-9mm}\caption{Functions $f(n/P)$ (left) and $g(n/P)$ (right) versus $n/P$.}\label{f-g}
\end{figure}

\begin{figure}[h]
\begin{center}$
\begin{array}{cc}
\hspace{-5mm}\includegraphics[scale=1]{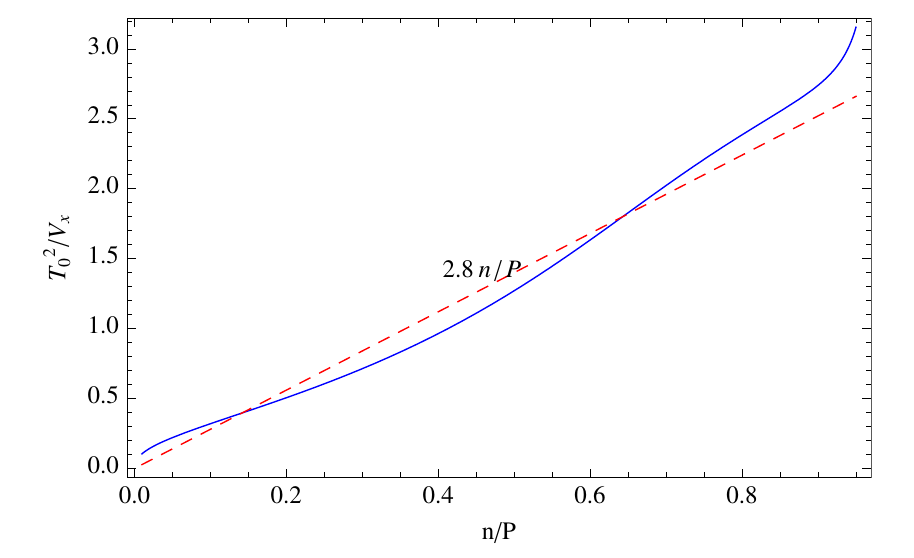} &
\includegraphics[scale=1]{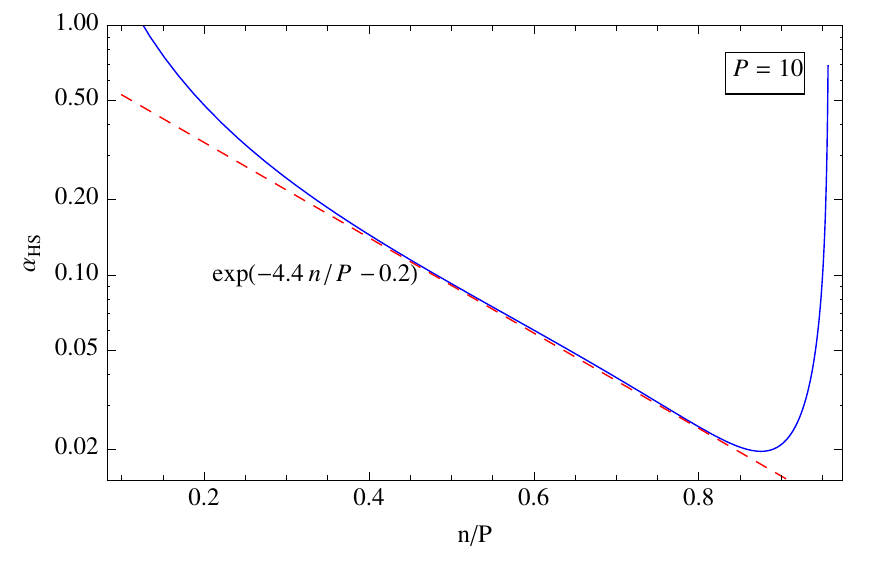}
\end{array}$
\end{center}
\vspace{-9mm}\caption{$\overline T_aT_a/V_X$ vacuum expectation values (left) and $\alpha_{\rm HS}$ (right) versus $n/P$. The dashed lines approximate those functions which are good enough in the range $0.2\lesssim n/P\lesssim 0.8$. }\label{TT-alpha}
\end{figure}

One can actually solve the above equations for $T_a\overline T_a/V_X$ and $\alpha_{\rm HS}$ to find the condition for a solution in a de Sitter vacuum as a function of $n/P$. We have solved them both analytically and numerically. Meta-stable de Sitter vacua only exist for the case $n/P<1$, not for $n/P>1$. The solutions are given as follows
\ba \frac{\overline TT}{V_X} &=& \frac{n/P}{1-n/P}\Big(1 + \big(9/14-n/P\big)f(n/P)\Big)\ ,\\ \frac{\alpha_{\rm HS}\cdot P}{2\pi} &=& \frac{1}{2}\frac{1-n/P}{n/P} \bigg(1+\frac{n/P}{1+g(n/P)}\Big(\frac{14}{9}+\frac{5}{9}(9/14-n/P)f(n/P)\Big)\bigg)^{-1}\ .\ea 
The functions $f(n/P)$ and $g(n/P)$ have relatively complicated, uninteresting forms, they are plotted in figure \ref{f-g}. We also plot $\overline TT/V_X$ and $\alpha_{\rm HS}$ versus $n/P$ in figure \ref{TT-alpha}.
To further simplify the future formulae we use approximations to the vacuum expectation values and $\alpha_{\rm HS}$ as a function of $n/P$, which are also given in the figure.

\paragraph*{Independent Numerical Calculation}
We have numerically checked that our formulae give the correct results as a cross-check that our methodology is robust. We give here an outline
of a simple set of examples so that the reader has an idea of the numbers that we obtain.

We choose a compact $G_2$ manifold with 2 moduli and volume 
\be V_X = \sum_{k=1}^3 c_k \prod_{i=1}^{2} s_i^{a_i^k}\ ,\ee with the following microscopic parameters \ba && c_1=1,\ c_2=1/3,\ c_3=1,\cr && a_1^1=7/6,\ a_2^1=7/6,\ a_1^2=1,\ a_2^2=4/3,\ a_1^3=1/3,\ a_2^2=2,\cr && N_1=N_2=1\, .\nn\ea
We scanned over all parameter space. The vev's of the moduli and scalar fields for different values of $n/P$ that we obtained are of order 
\ba \langle s_i\rangle &\approx& 10 -50 \ ,\\ \langle \overline TT/V_X\rangle &\approx& 1.1 - 0.3\ .\ea  There was good agreement with our formulae.
The fine-tuning of cosmological constant is done through $\lambda/A$ which take values between 0.1 and 100.


\section{Phenomenology}
In this section we will study some of the phenomenology of the $G_2$ vacua with emphasis on the soft SUSY breaking parameters. The values of the soft SUSY breaking parameters at the unification scale is determined in the standard way; the moduli and the $F$-terms are replaced by their vacuum expectation values in the supergravity Lagrangian. We then take the limit $m_{\rm pl}\rightarrow\infty$ while keeping $m_{3/2}$ fixed to obtain the soft supersymmetry breaking Lagrangian. Having determined soft parameters at the unification scale we will RG evolve them to the electroweak scale.

\subsection{Stabilized Volumes}
\begin{figure}[h]
\begin{center}$
\begin{array}{cc}
\hspace{-8mm}\includegraphics[scale=1]{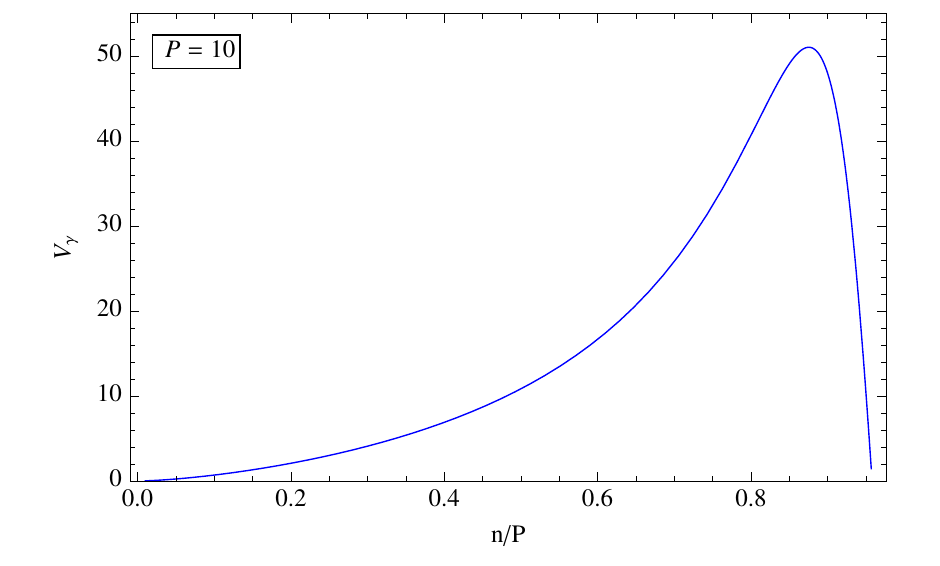} &
\hspace{-5mm}\includegraphics[scale=1]{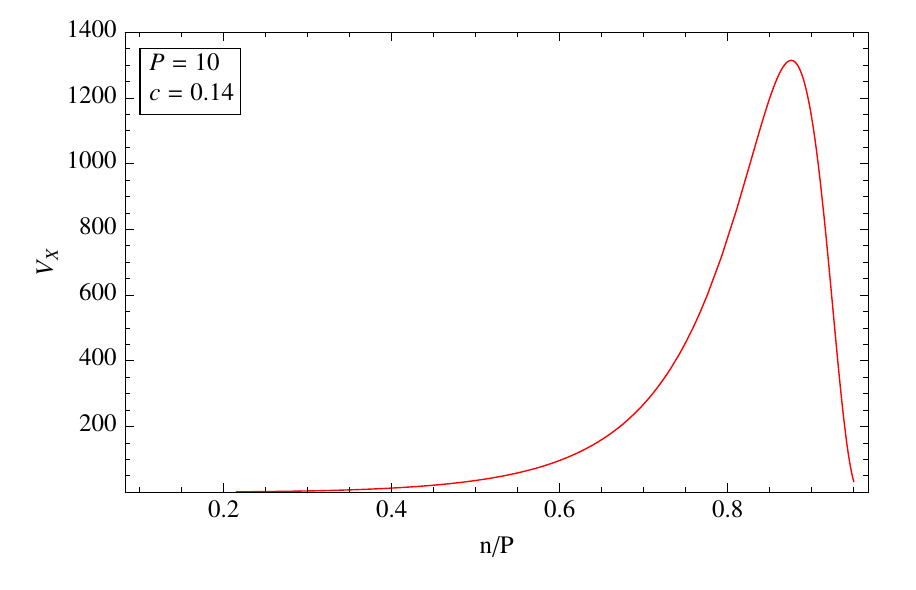}
\end{array}$
\end{center}
\vspace{-9mm}\caption{Volume of associative 3-cycle $V_{\gamma_3}$ (left) and volume of the compactification $V_X$ (right) versus $n/P$.}\label{volumes}
\end{figure}

There are basically two volumes which control the low scale phenomenology, the volume of the compactification manifold $V_X$ and the volume of associative 3 cycle $V_{\gamma_3}$. They are homogeneous functions of the moduli of degrees 7/3 and 1 respectively. Since we are interested in the weak coupling limit and $V_{\gamma_3} =
\alpha_{\rm HS}^{-1}$, we expect that
\be\label{V_X-V_gamma} V_X = \prod_{i=1}^{b_X^3} s_i^{a_i} = c\ \alpha_{\rm HS}^{-7/3}\ ,\qquad{\rm and}\qquad V_{\gamma_3} = \sum_{i=1}^{b_X^3} N_is_i = \alpha_{\rm HS}^{-1}\ .\ee 
In the above $c$ is a constant which plays a useful role \cite{Friedmann-Witten}. The first relation implies that the linear combination of moduli which determine the volume of
the 3-cycle $\gamma_3$ dominate the volume function. 
We have numerically checked that, for generic compactification manifolds, it is a good approximation for $V_X$.
Once the moduli get vacuum expectation values the volumes also become fixed. Phenomenological parameters of the low energy physics are functions of these volumes. They are plotted in figure \ref{volumes}.

The dimensionless parameter $c$ in \eqref{V_X-V_gamma} is 
\be c=\frac{V_X}{V_{\gamma_3}^{7/3}}\ , \ee 
and it is useful to have an estimate of its range of values. We combine the two relations in \eqref{V_X-V_gamma} in two extreme cases in order to get bounds on $c$. In one limit we take all $a_i$'s equal to $7/3N$ and in the other limit we take one of $a_i$'s, say $a_0$, much bigger that the rest. It gives the following bounds 
\be {\cal O}(10^{-2})\sim(N\bar N)^{-7/3} \lesssim c \lesssim \Big(\frac{3a_0}{7N_0}\Big)^{7/3}\sim{\cal O}(1)\ ,\ee where $N$ is the number of moduli and $\bar N$ is the geometric mean of $N_i$'s. Note that these bound are coming from microscopic properties of the compactification manifold $X$.

Similar bounds comes from phenomenological requirements. Loosely speaking, validity of the supergravity approximation requires 
\be V_X\gtrsim 1 .\ee On the other hand, in order to make $M$ theory (KK threshold) corrections to the parameters of the four dimensional physics negligible at and below $m_{\rm GUT}$, we demand 11 dimensional Planck mass $m_{11}$ to satisfy 
\be m_{11}>m_{\rm GUT} \sim {\cal O}(10^{16}{\rm GeV}).\ee 
From dimensional reduction, one finds the following relation between the 4 and 11 dimensional Planck mass ($m_{\rm pl} = 2.43\times 10^{18}$ GeV) \be \frac{m_{\rm pl}}{m_{11}} = \left(\frac{V_X}{\pi}\right)^{1/2}\ .\ee 
Thus, one finds the following upper bound on $V_X$ \be V_X \leq \pi\left(\frac{m_{\rm pl}}{m_{\rm GUT}}\right)^2\ .\ee 
All things together, we need $V_X$ to satisfy the following constraint \be {\cal O}(1)\lesssim V_X\lesssim {\cal O}(10^4)\ .\ee 
Again, the lower bound comes from requiring validity of supergravity regime and the upper bound comes from requirement of having a high compactification scale. It is basically 
\be m_{\rm pl} \gtrsim m_{11}\gtrsim m_{\rm GUT}\ .\ee
To make these phenomenological constrains consistent, one demands \be {\cal O}(5) \lesssim V_{\gamma_3} \approx \alpha_{\rm GUT}^{-1} \lesssim {\cal O}(100) \ .\ee
As an example, for the case of the 3-cycle supporting the Standard Model gauge group, one finds the actual value at which the volume $V_X$ must be stabilized can be  determined from the GUT scale
($\alpha_{\rm GUT}^{-1} = V_{\gamma_3}$ )
\be c= \frac{V_X}{V_{\gamma_3}^{7/3}} = \frac{1}{4\pi}\alpha_{\rm GUT}^3\Big(\frac{m_{pl}}{m_{\rm GUT}}\Big)^2 \approx 7.51\times 10^{-2}\Big(\frac{\alpha_{\rm GUT}}{0.04}\Big)^3\Big(\frac{2.2\times 10^{16}{\rm GeV}}{m_{\rm GUT}}\Big)^2  \lesssim {\cal O}(1)\ .\ee  

\subsection{Gravitino Mass}
\begin{figure}[h]
\begin{center}$
\begin{array}{cc}
\hspace{-5mm}\includegraphics[scale=1]{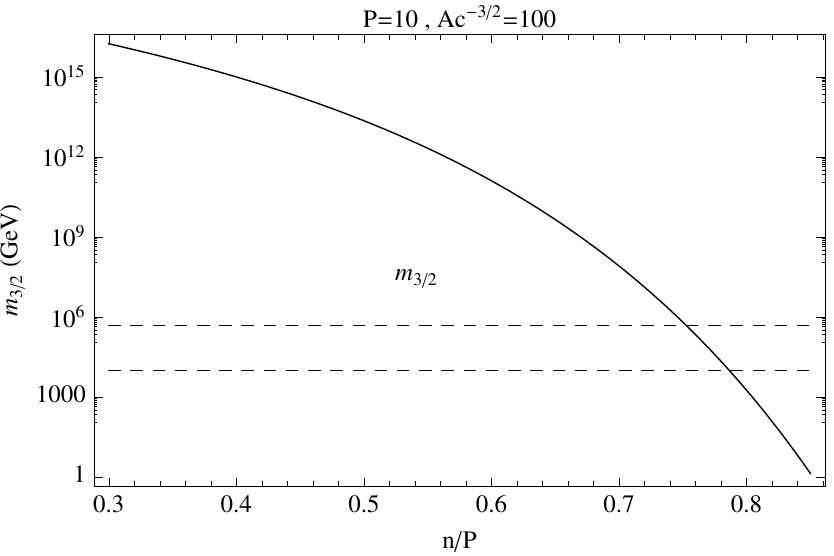} &
\includegraphics[scale=1]{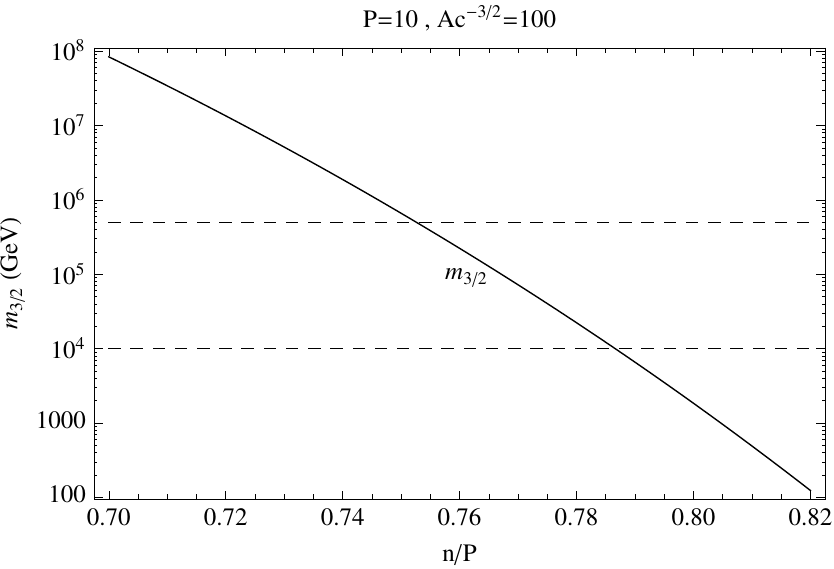}
\end{array}$
\end{center}
\vspace{-9mm}
\caption{Gravitino mass versus $n/P$. The lower dashed line is 10 TeV for reference to the cosmological BBN bound (30 TeV) and the upper dashed line is from non-thermal moduli dominated cosmology (500 TeV).}
\label{gravitino-plot}\end{figure}

The gravitino mass plays an important role in gravity mediated models of SUSY breaking and sets the typical mass scale in the supergravity Lagrangian. The bare gravitino mass is defined as 
\be\label{gravitino-mass} m_{3/2} = m_{\rm pl}\cdot e^{K/(2m_{\rm pl}^2)}\ |W/m_{\rm pl}^3|\ .\ee 
Having tuned the tree-level vacuum energy, we can determine gravitino mass using 
\be\begin{split}\label{gravitino-mass-explicite} m_{3/2} &= m_{\rm pl}\ \frac{e^{\sum_a\overline T_aT_a/2V_X}}{8\pi^{1/2} V_X^{3/2}}\ A(1-x)\, e^{-2\pi V_{\gamma_3}/ P}
\cr &= 0.035Ac^{-3/2}\,(1-n/P)\Big(1+0.46\frac{n/P}{1-n/P}Pe^{-4.4(n/P)}\Big) e^{-11/2(n/P)-7.66e^{4.4(n/P)}/P}m_{\rm pl}
\ . \end{split}\ee 
As can be seen, the gravitino mass is determined by four parameters, $n/P$, $P$, $A$ and $c$. The following plots in figure \ref{gravitino-plot} shows the behavior of $m_{3/2}$ versus $n/P$. 

\subsection{Embedding the MSSM}
The MSSM, being a gauge theory with charged chiral matter, resides on a 3-manifold (with orbifold and conical singularities) different from the one supporting the hidden sector. Since two 3-manifolds in a 7-dimensional manifold generically do not intersect, the supersymmetry breaking generated by strong gauge dynamics in the hidden sector is generically mediated to visible sector via Planck suppressed interactions {\it i.e.} it is gravity mediated.  In the analysis in the following sections, it is assumed that there is a GUT gauge group in the visible sector which is broken to the SM gauge group by background gauge fields (Wilson lines), following Witten \cite{splitting}.

The full low energy supergravity theory of the visible and hidden sectors at the compactification scale (unification scale) is defined by the following K\"ahler, superpotential and gauge kinetic functions
\ba K &=& K + K^{U_1} + K^{\rm VS} = -3\ln\big(4\pi^{1/3}V_X\big) + \frac{\sum_a\overline T_aT_a}{V_X} + K^{VS}_{\bar\alpha\beta}(s_i)\bar\Phi^{\bar\alpha}\Phi^\beta + \big(Z(s_i)H_uH_d+h.c.\big),\quad \\ W &=& W^{\rm HS} + W^{U_1} + W^{\rm VS} = A\,e^{2\pi if/P} + \lambda\,T_1T_2T_3\,e^{2\pi i\tilde f} + Y^{VS}_{\alpha\beta\gamma}\Phi^\alpha\Phi^\beta\Phi^\gamma, \\ \label{gauge-kinetic-functions} f &=& \sum_{i=1}^{b_X^3}N^{\rm HS}_iz_i\ ,\quad \tilde f = \sum_{i=1}^{b_X^3}\tilde N_iz_i\ ,\quad f^{VS} = \sum_{i=1}^{b_X^3}N^{\rm VS}_iz_i\ , \ea where $\Phi$'s are chiral matter fields, $K_{\bar\alpha\beta}$ is the K\"ahler metric and $Y_{\alpha\beta\gamma}$ are un-normalized Yukawa couplings in the visible sector. Evidently, there is no  bilinear (mass)  $\mu$ term in the superpotential. It can be forbidden by a discrete symmetry which ensures doublet-triplet splitting \cite{splitting}. This term can be generated by non-renormalizable terms in the K\"ahler potential (via Giudice-Masiero mechanism \cite{Giudice:1988yz}) or in the  superpotential. 

The K\"ahler metric $K^{\rm VS}_{\bar\alpha\beta}$ for the visible sector matter is assumed to be approximately flavor-universal and given by\be K^{\rm VS}_{\bar\alpha\beta} = \frac{1}{V_X}\Big(\delta_{\bar\alpha\beta}+\gamma_{\bar\alpha\beta}\frac{\sum_a\overline T_aT_a}{3V_X}\Big).\ee This can be ensured by $U(1)$ symmetries
associated to the vanishing 2-cycles at the location of the matter multiplets \cite{Bourjaily:2009vf}.
Higher order flavor-non-diagonal contributions to the K\"ahler potential can come from gravitational couplings between the scalars $T_a$ and the matter fields $\Phi_\alpha$ in the visible sector, but these are expected to be small due to the $U(1)$ symmetry.
Hence, the K\"ahler metric of the visible sector is diagonal and the K\"ahler potential can be written as \be K^{\rm VS} = \tilde K^{\rm VS}\delta_{\bar\alpha\beta}\bar\Phi^{\bar\alpha}\Phi^{\beta}.\ee

The gauge kinetic function in the visible sector is an integer linear combination of the moduli fields. The integers $N_i^{\rm VS}$ is fixed by homology class of the 3-cycle supporting the observable sector gauge and matter fields.

The Yukawa couplings arise from membrane instantons which connect singularities where chiral superfields are supported and given by \be Y_{\alpha\beta\gamma} =c_{\alpha\beta\gamma}e^{2\pi i\sum_il^i_{\alpha\beta\gamma}z_i}\ ,\ee where $c\sim{\cal O}(1)$ and $l^i\in{\mathbb Z}$. 
Thus, due to exponential dependence on the moduli it is natural to obtain a hierarchical structure of Yukawa couplings \cite{Acharya-Witten, AcharyaYukawa}.

For simplicity, here we assume an observable sector which is precisely the MSSM, though $G_2$-manifolds with additional matter can be treated
in exactly the same way.
Having all the moduli fixed, the $F$-terms can be explicitly computed in terms of the constants which determine the moduli potential e.g. $n/P$. 
Given a particular $G_2$-manifold and a particular set of microscopic constants one can obtain a particular point in the MSSM parameter space. This defines a particular ``$G_2$-MSSM''.

\subsection{Physics at the Unification Scale}
\subsubsection{Gaugino Masses}
In general three different sources contribute to the gaugino masses $m_{1/2}$, the tree-level and anomaly mediation contribution as well as threshold corrections. They can be either universal or non-universal. The total gaugino  masses can be read as 
\be m_a(m_{\rm GUT}) = m_{\rm 1/2}^{\rm tree} + m_{1/2}^{\rm threshold} + m_a^{\rm AMSB}\ .\ee
As will be seen, in $G_2$ vacua the tree-level gaugino masses are suppressed relative to gravitino mass. This is due to the fact that the $F$-terms of the moduli are suppressed compared to those of the matter fields and the gauge coupling functions are independent of the $T_a$. This also happens
in \cite{G2-MSSM}. Because the tree-level contributions to $m_{1/2}$ are smaller than $m_{3/2}$, one has to carefully consider the anomaly mediation and threshold
corrections which could therefore be comparable to the tree level terms.
In the following these contributions are evaluated at the unification scale. 

\begin{figure}[h]
\begin{center}$
\begin{array}{cc}
\hspace{-5mm}\includegraphics[scale=1]{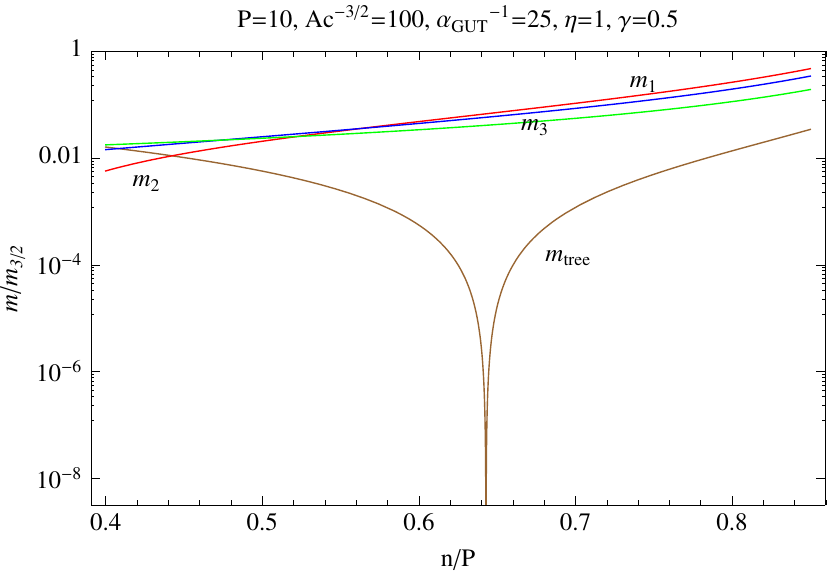} &
\includegraphics[scale=1]{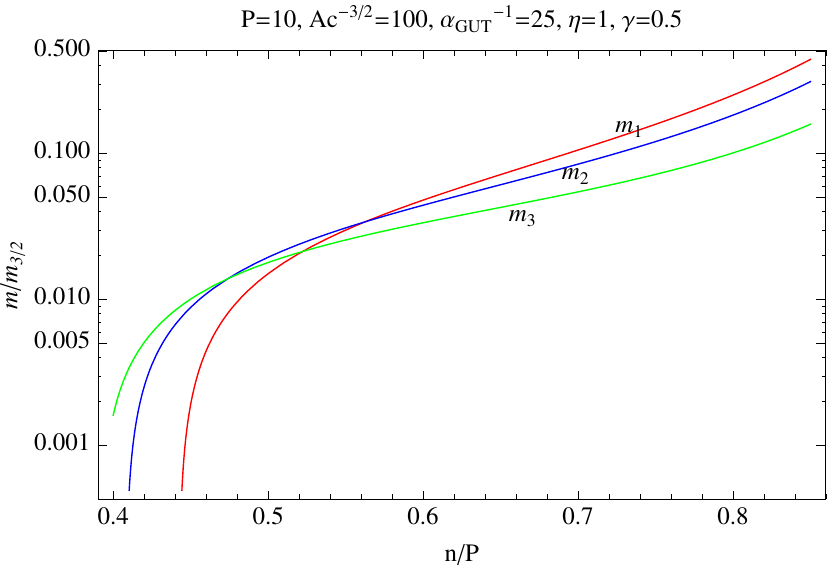}
\end{array}$
\end{center}
\vspace{-9mm}\caption{This plot shows gaugino to gravitino mass ratios. On the left we have separately plotted the tree level and AMSB contributions. On the right, they are summed up. It is clear that for all values of $n/P$ where expansion makes sense, this ratio is smaller than one.}
\label{gaugino-plot}\end{figure}

\paragraph{Tree-level contribution}The universal (having assumed gauge unification) tree level contribution to the gaugino masses can be computed as \cite{Brignole-Ibanez-Munoz, Nilles:1983ge}
\be m_{1/2}^{\rm tree} = \frac{1}{2i{\rm Im}f_{\rm VS}}\,e^{K/2}F^i\partial_if_{\rm VS}\ ,\ee where the tree-level visible sector gauge kinetic function is given in \eqref{gauge-kinetic-functions}. As mentioned earlier, we assume a GUT group broken to the SM by a discrete choice of Wilson lines. Thus, at the unification scale the gauge kinetic function of 3 gauge groups in the SM are the same. This implies that the tree level gaugino masses are also unified. 
We find 
\be m_{1/2}^{\rm tree} = -0.13\zeta\frac{n/P}{1-n/P}Pe^{-4.4(n/P)}\bigg(\frac{2}{3}\frac{1}{1+2.8(n/P)}\frac{1}{1-n/P} - \frac{5}{21} - \frac{5}{98}\frac{1}{1-n/P}\bigg) m_{3/2}\ ,\ee 
where we have dropped the gaugino phases $(N_it_i)/P$.
The result from analytical calculation is plotted in figure \ref{gaugino-plot}. As can be seen, there is a funnel at $n/P\sim 9/14$ where the $F$-terms vanishes at tree level.

\paragraph{Threshold Correction} The threshold corrections to the gauge kinetic function from the Kaluza-Klein modes are computed in \cite{Friedmann-Witten} 
\be\label{alpha_GUT} \alpha_{\rm GUT}^{-1} = {\bf Im}(f^{VS}) + \delta\ ,\ee 
where 
\be \delta = \frac{5}{2\pi}{\cal T}_\omega  =\frac{5}{2\pi}\ln\Big(4\sin^2\frac{5\pi\omega}{q}\Big) \ .\ee 
In the above, ${\cal T}_\omega$ is a topological invariant and $q$ and $\omega$ are integers determined by the topology of $X_7$. Then, we invert \eqref{alpha_GUT} to find
\be\label{eta} {\bf Im}(f^{VS}) = \alpha_{\rm GUT}^{-1} - \delta = \eta\ \alpha_{\rm GUT}^{-1}\ ,\ee
where now
\be \eta \equiv 1-\alpha_{\rm GUT}\delta\ .\ee
Summing up both contributions we fine \be m_{1/2}^{\rm tree+threshold} = \eta \frac{1}{2i{\rm Im}f_{\rm VS}}\,e^{K/2}F^i\partial_if_{\rm VS}\ .\ee

\paragraph{Anomaly Mediation} Due to the substantial suppression of gaugino masses, it makes sense to take into account the anomaly mediated corrections \cite{AMSB}. These contributions are given by the following general formula at the unification scale \cite{AMSB1}
\be m_a^{\rm anomaly} = -\frac{\alpha_{\rm GUT}}{4\pi}\bigg[-\big(3C_a-{\cal C}_a\big)e^{K/2}\overline W + \big(C_a-{\cal C}_a\big)e^{K/2}\sum_IF^IK_I + 2\,{\cal C}_a e^{K/2}\sum_IF^I\partial_I\ln\tilde K^{\rm VS}\bigg],\ee where $C_a=0,2,3$ and ${\cal C}_a\equiv\sum_\alpha C_a^\alpha=33/5,7,6$ are the quadratic Casimirs of the $a^{th}$ gauge group. In our case it can be written in a more compact form as \be m_a^{\rm anomaly} = -\frac{\alpha_{\rm GUT}}{4\pi}\bigg[-\big(3C_a-{\cal C}_a\big)+\sum_IG^IK^{{\rm HS}+U_1}_I\Big(C_a-{\cal C}_a\big(1-2c_I\big)\Big)\bigg]e^{K^{{\rm HS}+U_1}/2}\overline W,\ee where the parameter $c_I$ is defined as the following ratio \be c_I=\frac{\partial_I\ln\tilde K^{\rm VS}}{K^{{\rm HS}+U_1}_I}\ .\ee 
At the minimum of the potential one obtains 
\ba\label{anomaly-at-minimum} \sum_{i=1}^{b_X^3} G^iK_i &=& \bigg[\frac{14}{3}\frac{1}{1-n/P}-\Big(\frac{5}{3}+\frac{15}{14}\frac{1}{1-n/P}\Big)\Big(1+2.8(n/P)\Big)\bigg]0.46\,\frac{n/P}{1-n/P}Pe^{-4.4(n/P)}  ,\\ \sum_{a=1}^3 G^aK_a &=& -3\frac{n/P}{1-n/P}+\frac{7}{9}\frac{23.5(n/P)^2}{1+2.8(n/P)} - 8.4(n/P)\bigg(\frac{4}{3}-\frac{7}{9}\frac{1}{1+2.8(n/P)}\frac{3-n/P}{1-n/P}\bigg)\cr &&-\bigg[\frac{-8/9 n/P}{1-n/P}+8.4(n/P)\bigg(\frac{5}{9}-\frac{7}{9}\frac{1}{1+2.4(n/P)}\frac{4n/P}{1-n/P}\bigg)\bigg]0.46\,\frac{n/P}{1-n/P}Pe^{-4.4(n/P)}, \nn\\\ea
Furthermore, one can see that for different moduli $s_i$ and $T_a$ \ba c_i &=& \frac{1+\frac{1}{1+\frac{1}{\gamma\frac{\sum_a\overline T_aT_a}{3V_X}}}}{3+\frac{\sum_a\overline T_aT_a}{V_X}} \approx \frac{1+\frac{1}{1+\frac{1}{2.8\gamma(n/P)}}}{3+8.4(n/P)}, \\ c_a &=& \frac{\gamma/3}{1+\gamma\frac{\sum_a\overline T_aT_a}{3V_X}} \approx \frac{\gamma/3}{1+2.8 \gamma(n/P)}\ .\ea
Putting all pieces together, one finds 
\be m_a^{\rm anomaly} = -\frac{\alpha_{\rm GUT}}{4\pi}\bigg[-\big(3C_a-{\cal C}_a\big) + \sum_{i=1}^{b_X^3} G^iK_i\big(C_a-{\cal C}_a(1-2c_i)\big) + \sum_{a=1}^{3} G^aK_a\big(C_a-{\cal C}_a(1-2c_a)\big)\bigg]m_{3/2}\ .\ee
Here we have also dropped the phase factor which is identical to the one in the tree level contribution. The anomaly mediated contributions are numerically comparable to the tree-level terms but are not universal. Thus, the gaugino masses are non-universal.
The anomaly contributions are also plotted in figure \ref{gaugino-plot}. 

\begin{figure}[h]
\begin{center}$
\begin{array}{cc}
\hspace{-5mm}\includegraphics[scale=1]{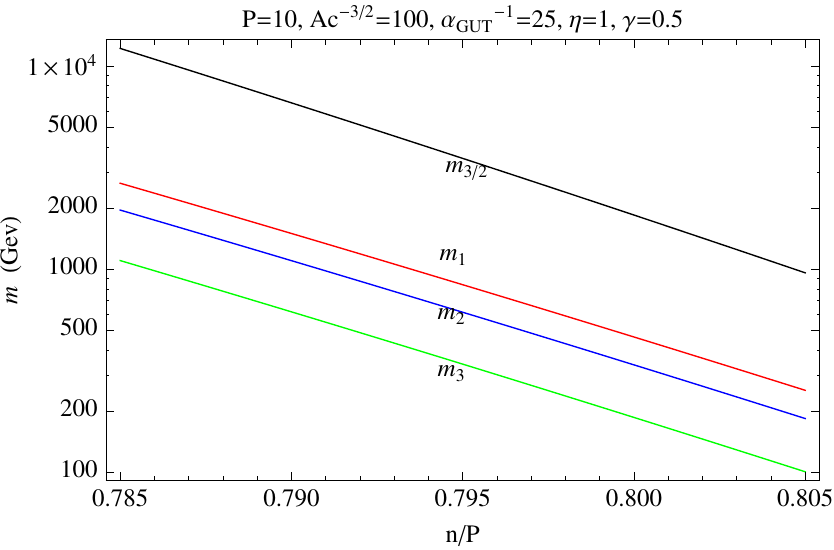} & \includegraphics[scale=1]{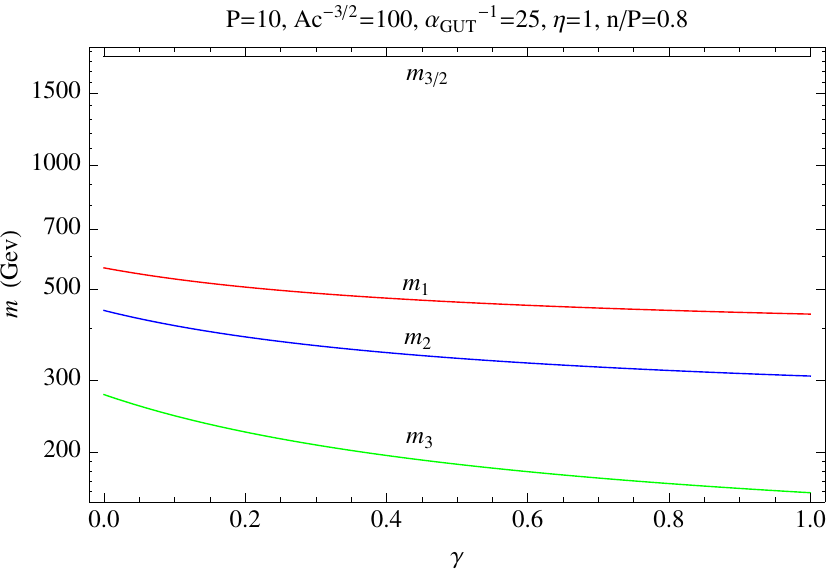} \cr
\includegraphics[scale=1]{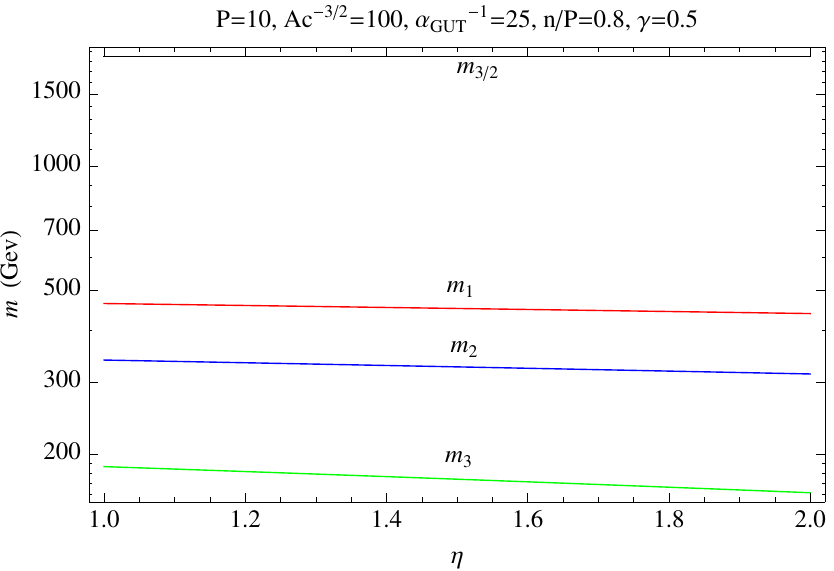} & \includegraphics[scale=1]{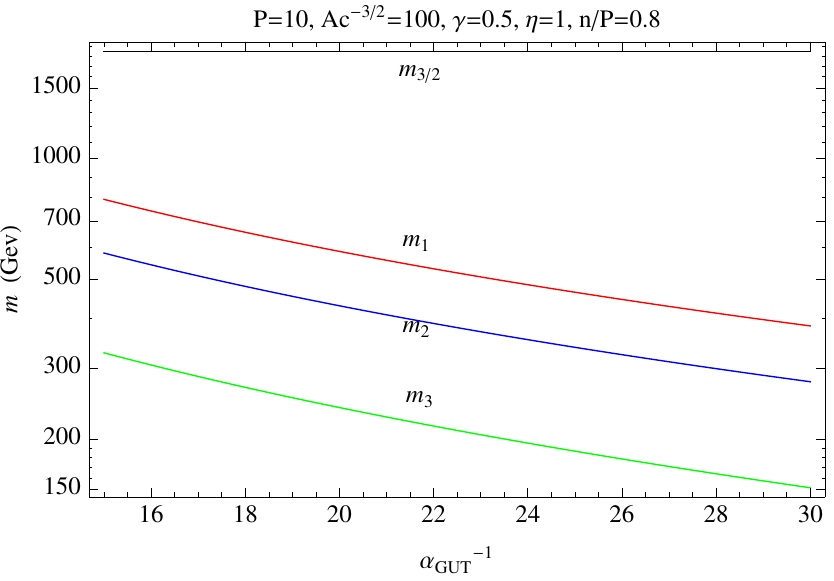}
\end{array}$
\end{center}
\vspace{-9mm}\caption{Masses versus $n/P$ (upper left). Masses for fixed $n/P$ versus $\gamma$ (upper right), $\eta$ (lower left) and $\alpha_{\rm GUT}$ (lower right).}\label{gaugino-gravitino-plot}  
\end{figure}

In summary, the full gaugino mass at the unification scale depends on the parameters $n/P$, $P$, $A$, $c$, ${\bf Im}f^{\rm VS}$, $\gamma$ and $\eta$ (or equivalently $\delta$). The ratio of the gaugino masses to the gravitino mass just depends on ${\bf Im}f^{\rm VS}$, $\gamma$ and $\eta$ which are subject to the constraint \eqref{eta}. The total gaugino masses and the gravitino mass are plotted in figure \ref{gaugino-gravitino-plot} in a restricted range of parameters.

\subsubsection{Scalar Masses}
\begin{figure}[h]
\begin{center}$
\begin{array}{cc}
\hspace{-5mm}\includegraphics[scale=1]{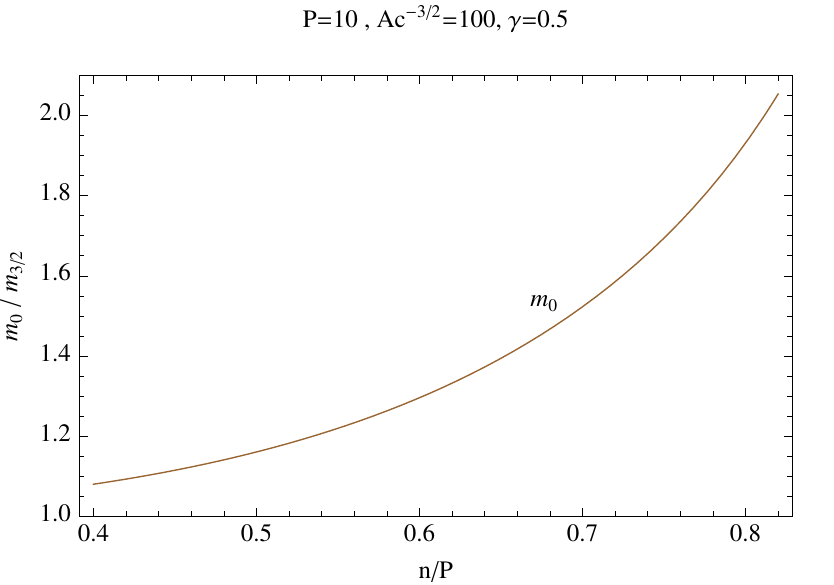} &
\includegraphics[scale=1]{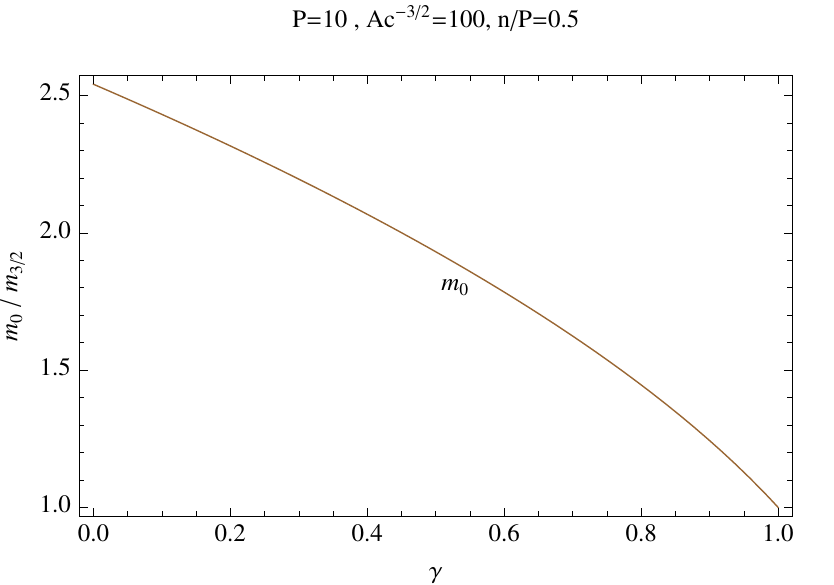}
\end{array}$
\end{center}
\vspace{-9mm}\caption{Scalar mass to gravitino mass ratio versus $n/P$ (left) and $\gamma$ (right).}\label{scalar-plot}
\end{figure}

The general formula for the un-normalized scalar masses is \cite{Brignole-Ibanez-Munoz} \be m_{\bar\alpha\beta}'^2 = \big(m_{3/2}^2+V_0\big)K^{\rm VS}_{\bar\alpha\beta} - \sum_{I,\bar J}e^KF^{\bar I}\big(\partial_{\bar I}\partial_J K^{\rm VS}_{\bar\alpha\beta} - \partial_{\bar I}K^{\rm VS}_{\bar\alpha\gamma}K^{{\rm VS},\gamma\bar\delta}\partial_JK^{\rm VS}_{\bar\delta\beta}\big)F^J.\ee One has to canonically normalize the visible sector K\"ahler potential by introducing unitary normalization matrices. The scalar masses generically have a flavor diagonal contribution of order the gravitino mass and a flavor non-diagonal contribution proportional to gravitino mass. The latter is hard to calculate and it is argued above and in \cite{G2-MSSM} they are small compared to the former. In the limit of a flavor diagonal K\"ahler metric and tuned cosmological costant, the masses for the canonically normalized scalars are 
\be m^2_{\alpha\bar\beta} = \Big(m_{3/2}^2 - \sum_{I,\bar J}e^{K^{X+U_1}}F^IF^{\bar J}\partial_I\partial_{\bar J} \ln\tilde K^{\rm VS}\Big)\delta_{\alpha\bar\beta}\ .\ee 
In the minimum of the potential, one obtains the following expression in the leading order 
\be m_0 = \bigg[1 + \frac{1-\gamma}{3}\bigg(18.2(n/P) + 3\frac{n/P}{1-n/P} - \frac{6.5(n/P)}{1+2.8(n/P)} \frac{3-n/P}{1-n/P} - \frac{18.2(n/P)^2}{1+2.8(n/P)}\bigg)\bigg]^{1/2}m_{3/2}\ . \ee 
The anomaly contributions to the scalar masses are suppressed relative to the order gravitino mass terms and we will thus neglect such contributions. In figure \ref{scalar-plot} the scalar masses are plotted against $n/P$ and $\gamma$.

\subsubsection{Trilinear Couplings} 
The un-normalized trilinear couplings are given by the following general expression \cite{Brignole-Ibanez-Munoz} 
\be A'_{\alpha\beta\gamma} = e^{K/2}\frac{\overline W}{|W|}\sum_IF^I\bigg[K_IY'_{\alpha\beta\gamma} + \partial_IY'_{\alpha\beta\gamma} - \Big(K^{\rm VS,\delta\bar\rho}\partial_IK^{\rm VS}_{\bar\rho\alpha}Y'_{\delta\beta\gamma} + K^{\rm VS,\delta\bar\rho}\partial_IK^{\rm VS}_{\bar\rho\beta}Y'_{\alpha\delta\gamma} + K^{\rm VS,\delta\bar\rho}\partial_IK^{\rm VS}_{\bar\rho\gamma}Y'_{\delta\beta\delta} \Big)\bigg].\ee 
In the case of a diagonal K\"ahler metric in the visible sector and under assumptions we made in scalar masses, the normalized trilinear couplings are proportional to the physical Yukawa couplings $Y_{\alpha\beta\gamma}$ as 
\be A_{\alpha\beta\gamma} = Y_{\alpha\beta\gamma}e^{K^{X+U_1}/2}\overline W\sum_I G^I\Big[K_I+\partial_I\ln Y'_{\alpha\beta\gamma} - \partial_I \ln \big(\tilde K^{\rm VS}_{\alpha}\tilde K^{\rm VS}_{\beta}\tilde K^{\rm VS}_{\gamma}\big)\Big]. \ee As mentioned earlier, $Y'_{\alpha\beta\gamma}$ are the un-normalized Yukawa couplings which arise from Euclidean membranes wrapping associative 3-cycles and connect isolated conical singularities. Due to the approximate flavor-diagonal nature of the K\"ahler potential, one finds \be Y_{\alpha\beta\gamma} = e^{K^{({\rm HS}+U_1)/2}}|Y'_{\alpha\beta\gamma}|\big(\tilde K^{\rm VS}_{\alpha}\tilde K^{\rm VS}_{\beta}\tilde K^{\rm VS}_{\gamma}\big)^{-1/2} = e^{K^{({\rm HS}+U_1)/2}}|c_{\alpha\beta\gamma}|e^{-2\pi\sum_il^{\alpha\beta\gamma}_is_i}\big(\tilde K^{\rm VS}_{\alpha}\tilde K^{\rm VS}_{\beta}\tilde K^{\rm VS}_{\gamma}\big)^{-1/2}\ .\ee

Having done some algebra, one can show that, at the minimum of the potential, the trilinear coupling at tree-level can be read as 
\be A_{\alpha\beta\gamma} = Y_{\alpha\beta\gamma}\Bigg[\sum_{i=1}^{b_X^3} G^iK_i\bigg(1-\frac{1+2.8(n/P)(\gamma-1)}{1+2.8(n/P)}\bigg) + \gamma\sum_{a=1}^3G^aK_a + \sum_{i=1}^{b_X^3}G^i\partial_i\ln Y'_{\alpha\beta\gamma}\Bigg]m_{3/2}\ . \ee 
The first two terms are the same as in \eqref{anomaly-at-minimum} and the last term is the following
\be \sum_{i=1}^{b_X^3} G^i\partial_i\ln Y'_{\alpha\beta\gamma} = \frac{n/P}{1-n/P}\alpha_{\rm HS}\cdot P\bigg(\frac{14/3}{1-n/P}\frac{1}{1+2.8(n/P)} - \frac{5}{3} - \frac{15/14}{1-n/P}\bigg) V_{{\gamma_3}},\ee where $V{{\gamma_3}}$ is the volume of the 3-cycle connecting co-dimension 7 singularities where the chiral multiplets are localized. Generically, the trilinear couplings are the same order of gravitino mass.

\subsubsection{$\mu$ and $B\mu$ Terms}
Having set the bilinear term in the superpotential to zero, the general expressions for the normalized $\mu$ and $B\mu$ are given by \cite{Brignole-Ibanez-Munoz} 
\ba \mu &=& \Big(m_{3/2}Z-e^{K/2}\sum_I F^I\partial_I Z\Big)\big(K^{\rm VS}_{H_u}K^{\rm VS}_{H_d}\big)^{-1/2},\\ B\mu &=& \bigg[\big(2m_{3/2}^2+V_0\big)Z - m_{3/2}e^{K/2}\sum_{\bar I} F^{\bar I}\partial_{\bar I} Z + m_{3/2}e^{K/2}\sum_I F^I\Big(\partial_I Z-Z\partial_I\ln \big(K^{\rm VS}_{H_u}K^{\rm VS}_{H_d}\big)\Big) \cr && -e^K\sum_{\bar I,J}F^{\bar I}F^J\Big(\partial_{\bar I}\partial_J Z - \partial_{\bar I}Z\partial_J\ln\big(K^{\rm VS}_{H_u}K^{\rm VS}_{H_d}\big) \Big)\bigg]\big(K^{\rm VS}_{H_u}K^{\rm VS}_{H_d}\big)^{-1/2}\ .\ea The Higgs bilinear $Z$ in the K\"ahler potential is treated in detail in \cite{Mupaper}. In our study we will parametrize the $\mu$ and $B\mu$ terms as follows \ba \mu &=& Z^\mu m_{3/2},\\ B\mu &=& Z^{B\mu}m_{3/2}^2\ ,\ea 
where $Z^\mu$ and $Z^{B\mu}$ are ${\cal O}(1)$ parameters.

\subsection{Physics at the Electroweak Scale}
In the previous subsection, all of the soft SUSY breaking parameters were computed at the unification scale. In order to connect to low energy physics and get the MSSM spectrum at the electroweak scale we need to RG evolve the masses and couplings from the unification scale down to the electroweak scale. The detailed discussion of renormalizing the masses and couplings down to electroweak scale is given in \cite{G2-MSSM}. 

Due to the anomaly mediated contribution, the gaugino masses are sensitive to the value of $\alpha_{\rm GUT}$. However, the value of $\alpha_{\rm GUT}$ is only determined once we know the exact spectrum and run the gauge coupling up to the GUT scale. 

Although at the unification scale squarks and sleptons have approximately flavor universal masses, it turns out that RG evolution down to the electroweak scale increases the masses of the $1^{st}$ and $2^{nd}$ generation squark and sleptons and decreases the mass of the $3^{rd}$ generation. The former increase is mostly proportional to the gaugino masses and thus the effect is small. The latter decrease is due to large Yukawa couplings and is caused by the trilinear parameters. In particular, the right handed stop is typically the lightest of all sfermions. However, the $\mu$ parameter which determines the masses of the Higgsinos does not change much due to the standard non-renormalization theorems. 

As in \cite{G2-MSSM}, it is convenient to work with an effective theory with sfermions and Higgsinos integrated out at the decoupling scale $m_{\rm dec}\sim{\cal O}(m_{3/2})$. The low energy effective theory below this scale is the Standard Model plus gauginos. To take into account the threshold effects of integrated out fields, we exploit the match and run procedure.  In the rest of this subsection, we study the evolution of gaugino masses.

\begin{figure}[h!]
 \begin{center}$
 \begin{array}{cc}
 \hspace{0mm}\includegraphics[angle=0,scale=1]{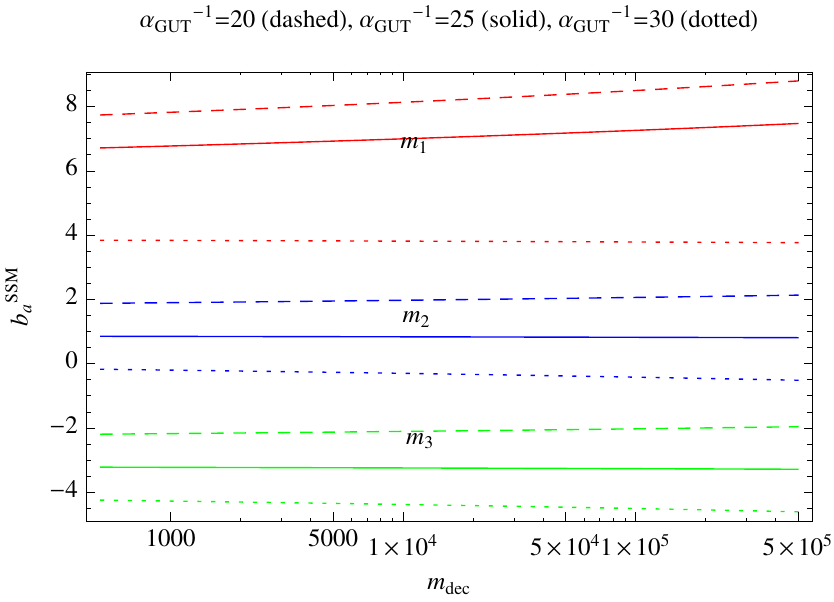} & \hspace{-2mm}\vspace{-7mm}\includegraphics[angle=0,scale=1]{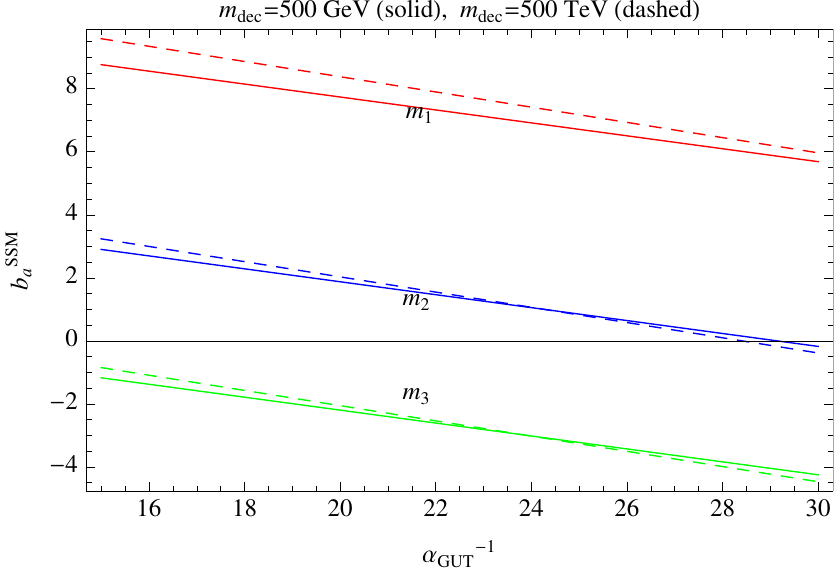}
 \end{array}$
 \end{center}
 \caption{The $\beta$ function coefficients for gauge couplings versus decoupling scale (left) and unified gauge coupling (right).}
 \label{beta-coefficients}
 \end{figure}

The gaugino masses at the electroweak scale at one-loop level can be related to those at the unification scale by a RG evolution factor  \be m_a(m_{\rm EW}) = K_a\ m_a(m_{\rm GUT})\ .\ee 
Whereas the running is done with tree-level matching at the decoupling scale, the RG evolution factor can be read as \be K_a = \frac{\alpha_a(m_{\rm dec})}{\alpha_{\rm GUT}}\bigg[\frac{\alpha_a(m_{\rm EW})}{\alpha_a(m_{\rm dec})}\bigg]^{\tilde b_a^{\rm{SM}+\tilde g}/b_a^{\rm{SM}+\tilde g}} .\ee The constants $\tilde b_a^{\rm{SM}+\tilde g}=0,-6,-9$ and $b_a^{\rm{SM}+\tilde g}=41/10,-11/6,-5$ are the 1-loop $\beta$ function coefficients of the gaugino masses and gauge coupling respectively.
At the decoupling scale, the gauge coupling is given by 
\be\label{alpha@dec} \alpha_a^{-1}(m_{\rm dec}) = \alpha^{-1}_{\rm GUT} + \frac{b_a^{\rm SSM}}{2\pi}\ln\Big(\frac{m_{\rm GUT}}{m_{\rm dec}}\Big), \ee 
and at the electroweak scale it is \be\label{alpha_EW} \alpha_a^{-1}(m_{\rm EW}) = \alpha^{-1}(m_{\rm dec}) + \frac{b_a^{\rm{SM}+\tilde g}}{2\pi}\ln\Big(\frac{m_{\rm dec}}{m_{\rm EW}}\Big). \ee $b_a^{\rm SSM}$ are the beta function coefficients for the model above the decoupling scale e.g.
for the MSSM with $\alpha_{GUT} \sim 1/25$,  $b_a^{\rm MSSM} = 66/10, 1, -3$.  Combining equations \eqref{alpha@dec} and \eqref{alpha_EW} in order to remove gauge coupling in the decoupling scale $\alpha(m_{dec})$ we obtain \be\label{alpha_EW-II} \alpha_a^{-1}(m_{\rm EW}) = \alpha^{-1}_{\rm GUT} + \frac{b_a^{\rm{SM}+\tilde g}}{2\pi}\ln\Big(\frac{m_{\rm dec}}{m_{\rm EW}}\Big) + \frac{b_a^{\rm SSM}}{2\pi}\ln\Big(\frac{m_{\rm GUT}}{m_{\rm dec}}\Big)\ . \ee 
In figure \ref{beta-coefficients} the $b_a^{\rm SSM}$ are plotted for different values of $m_{\rm dec}$ and $\alpha_{\rm GUT}$.

\begin{figure}[h!]
\begin{center}$
\begin{array}{cc}
\vspace{01mm}\hspace{-8mm}\includegraphics[angle=0,scale=1]{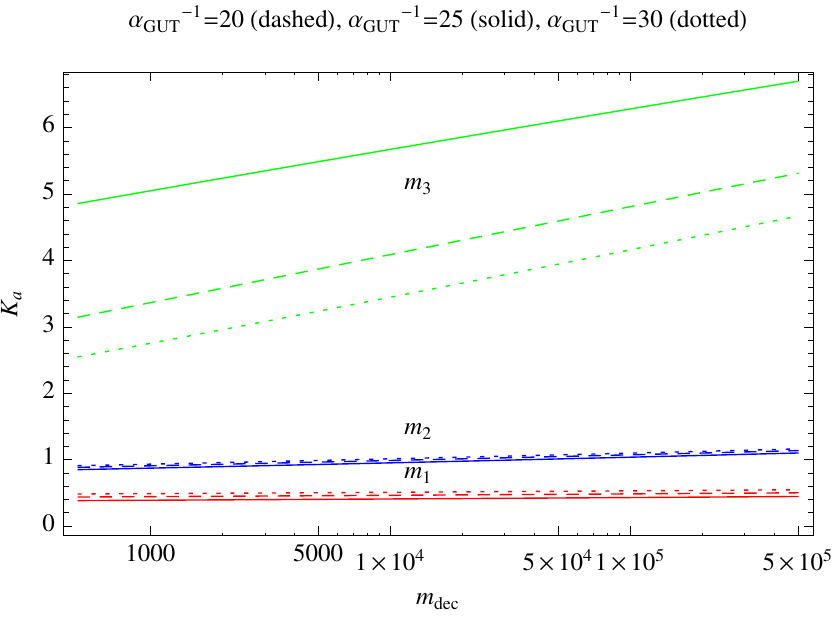} & \hspace{-5mm}\vspace{-8mm}\includegraphics[angle=0,scale=1]{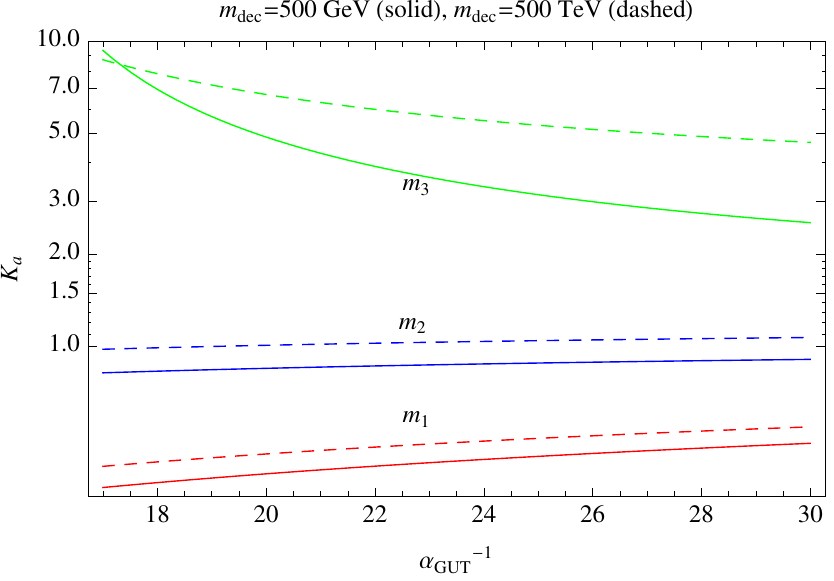}\end{array}$\end{center}
\caption{The RGE factors for gauge couplings versus decoupling scale (left) and unified gauge coupling (right) for the MSSM.}
\label{RGE-factors}
\end{figure}

Thus, the RGE factors can be determined for a given unified gauge coupling $\alpha_{GUT}$ and the decoupling scale $m_{dec}$ 
\be K_a = \frac{\left[1-\frac{b_a^{{\rm SM}+\tilde g}}{2\pi}\alpha_a(m_{\rm EW})\ln\left(\frac{m_{\rm dec}}{m_{\rm EW}}\right)\right]^{\tilde b_a^{\rm{SM}+\tilde g}/b_a^{\rm{SM}+\tilde g}}}{1+\frac{b_a^{\rm SSM}}{2\pi}\alpha_{\rm GUT}\ln\left(\frac{m_{\rm GUT}}{m_{\rm dec}}\right)}\ .\ee
They are plotted in figure \ref{RGE-factors} for different values of $m_{\rm dec}$ and $\alpha_{\rm GUT}$.

Moreover, weak threshold corrections must be added in order to obtain the gaugino pole masses. Since Higgsinos are as massive as gravitino, there is a substantial threshold contribution from the Higgs-Higgsino loops which must be taken into account when computing bino and wino masses. Under some circumstances these corrections can be large enough to alter the nature of the LSP.  The finite threshold correction is computed and given by \cite{Finite-1,Finite-2,Finite-3} 
\be \Delta m_{1,2}^{\rm finite} \approx \frac{\alpha_{1,2}}{4\pi}\, \mu\ .\ee 
This quantum correction can shift bino and wino masses up and down depending on the sign of $\mu$. Furthermore, the electroweak threshold corrections from gaugino-gauge boson loops must be included especially for the case of gluino 
 \be \Delta m_3 = \frac{3\alpha_3}{4\pi}\left(3\ln\left(\frac{m_{\rm EW}^2}{m_3^2}\right)+5\right)m_3\ .\ee 
 
\begin{figure}[h!]
\begin{center}$
\begin{array}{cc}
\hspace{-7mm}\includegraphics[scale=1]{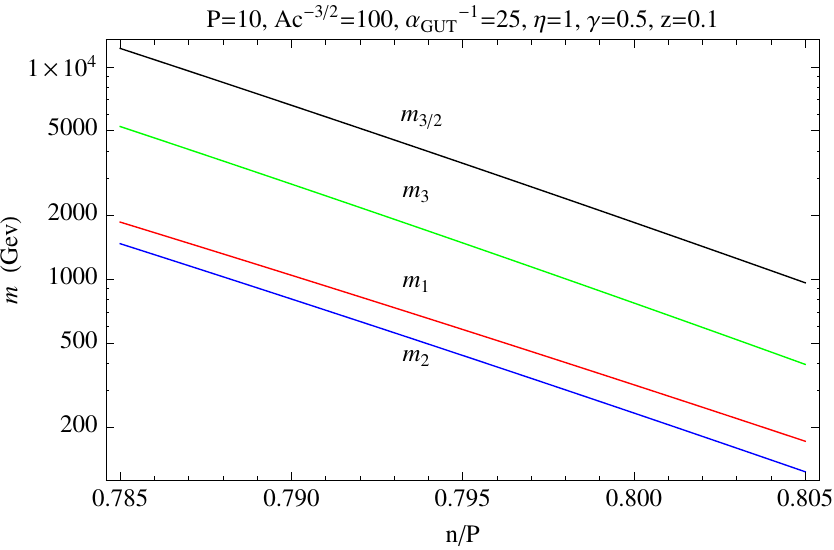} &
\end{array}$
\end{center}
\vspace{-9mm}\caption{Gaugino masses and gravitino mass versus $n/P$ at the electroweak scale.}\label{masses_EW-1}
\end{figure}

\begin{figure}[h!]
\begin{center}$
\begin{array}{cc}
\includegraphics[scale=1.]{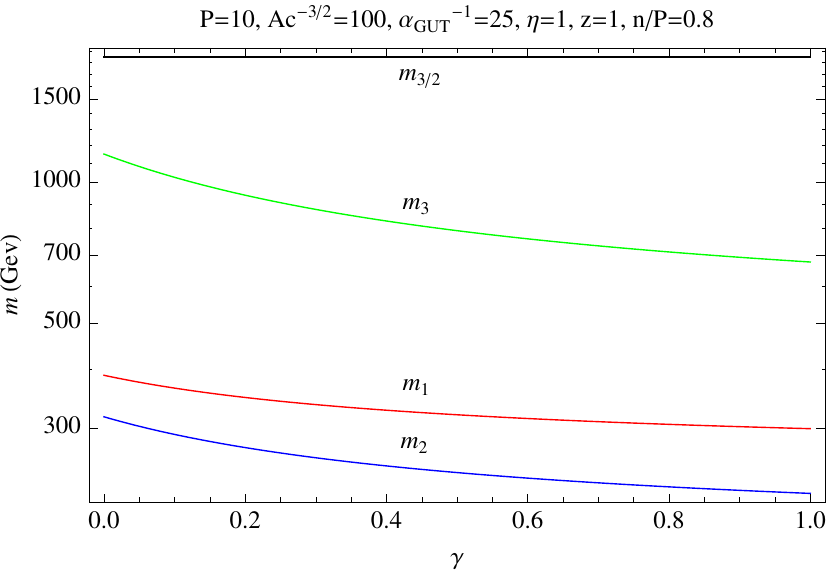} &
\includegraphics[scale=1.]{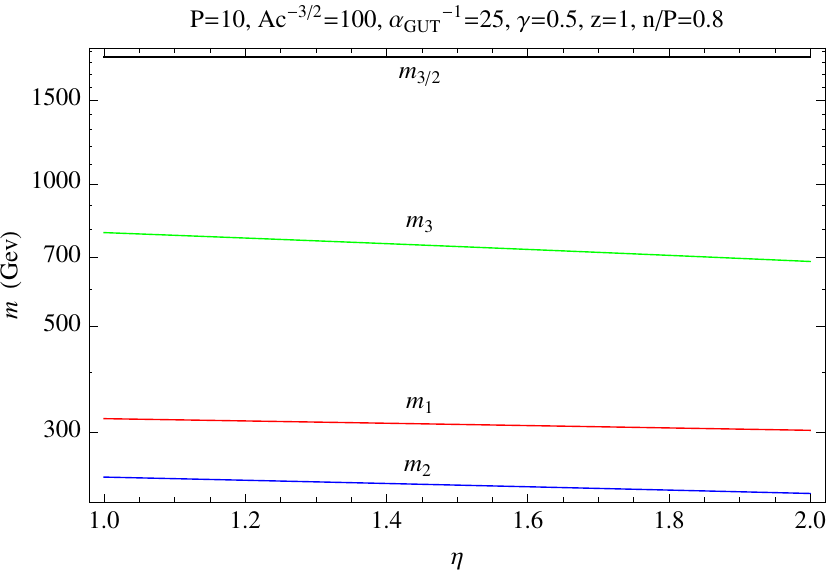} \cr 
\includegraphics[scale=1.]{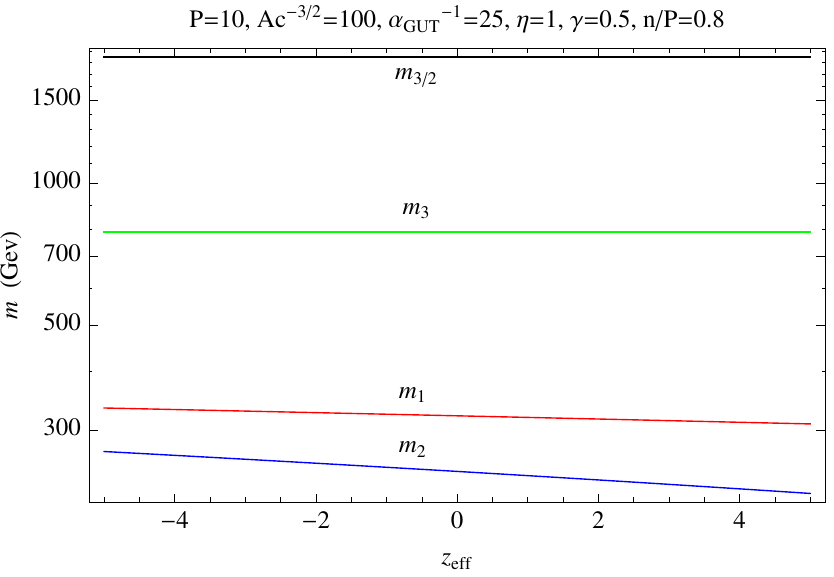} &
\includegraphics[scale=1.]{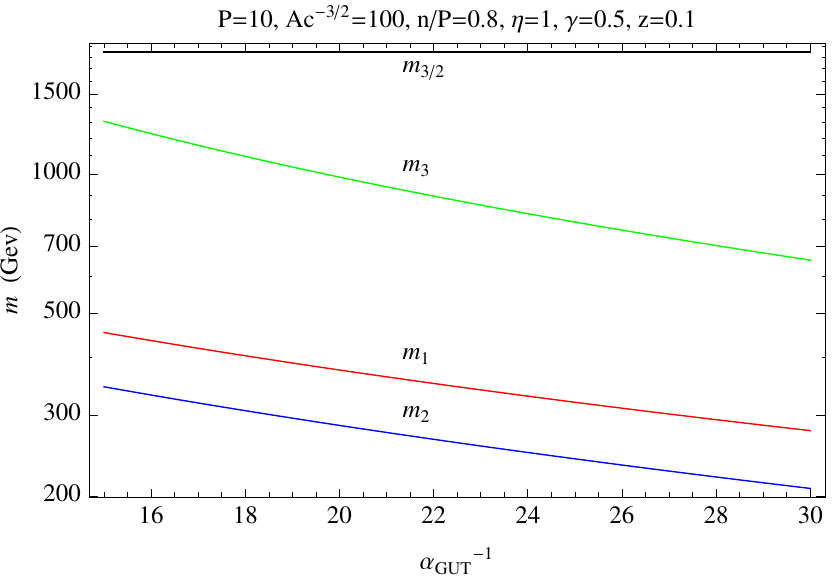} 
\end{array}$
\end{center}
\vspace{-9mm}\caption{Gaugino masses and gravitino mass versus $\gamma$ (upper left), $\eta$ (upper right), $z_{eff}$ (lower left) and $\alpha_{\rm GUT}$ (lower right)} \label{masses_EW-2}
\end{figure}

Taking all the various contributions into account, the gaugino masses  at the electroweak scale are \be m_a(m_{\rm EW}) = \Big(\eta\, m^{\rm tree} - m_a^{\rm AMSB}(\gamma)\Big) K_a + \Delta m^{\rm finite}(z_{eff})\ .\ee
In figures \ref{masses_EW-1} and \ref{masses_EW-2} gaugino masses at the electroweak scale are plotted. {\it Evidently, the wino is the LSP in these vacua}.

\section{Discussion}
We have considered a variation on the models considered in \cite{G2-MSSM} in which a cubic superpotential for matter
plays a significant role in moduli stabilization. This has allowed for a much wider range of gravitino masses than found
in \cite{G2-MSSM}, though for the phenomenologically viable window of $30 {\rm TeV} \leq m_{3/2} \leq 500 {\rm TeV}$ the mass spectrum
turns out to be practically identical to the models in \cite{G2-MSSM}. In particular, we find that the neutral wino
is the lightest supersymmetric particle. Note that a bino or Higgsino LSP would be inconsistent phenomenologically because
their annihilation cross-sections are not large enough when they are produced non-thermally from the moduli decays
before BBN \cite{non-thermal-miracle}. The LHC signals of these models include multi top quark production
from gluino pair production and decay as well as short track stubs from the chargino production and decay.

The models under consideration show that supergravity interactions can generate vacuum expectation values for charged matter fields
and thus break hidden sector gauge symmetries at a high scale, though the large vacuum expectation values which arise here really requires
an understanding of higher order corrections in the matter fields which we cannot compute at present and leave for future study.
It would also be interesting to consider different matter content in the observable sector e.g. $E_6$ representations
or the interesting single Higgs doublet models considered recently \cite{Ibe:2010ig}.

\acknowledgments
We would like to thank K. Bobkov, D. Feldman, G. Kane, E. Kuflik and P. Kumar for discussions.
The authors also thank the MCTP for hospitality where part of this project was done. MT appreciates the hospitality of the CERN theory division. 


\end{document}